\documentclass[myepj-spec,pdftex]{mySvjour}
\usepackage{graphicx}
\usepackage{hyperref}
\usepackage{color}
\usepackage{xspace}
\usepackage[square,sort&compress,numbers]{natbib}   
\usepackage{verbatim}
\usepackage{amsmath,amssymb,amsfonts} % Typical maths resource packages
\usepackage{tabularx}
\usepackage{multicol}
\usepackage{footnote}
\usepackage{listings}
\usepackage{multirow}
\usepackage{subfigure}
\usepackage[switch, modulo]{lineno}
\usepackage{lmodern,bm}                
\usepackage[T1]{sansmath} 
\SetMathAlphabet{\mathsfbf}{sans}{\sansmathencoding}{\sfdefault}{bx}{sl}
\usepackage{etoolbox}
\AtBeginEnvironment{sansmath}{}{}{}

\usepackage{lipsum}

\newcommand{\mW}{$m_{W}$}
\newcommand{\pTl}{$p_T^{\ell}$}
\newcommand{\mT}{$m_T$}
\newcommand{\pTnu}{$p_T^{\nu}$}

\definecolor{darkblue1}{rgb}{0,0,.2}
\definecolor{darkblue}{rgb}{0,0,.2}
\definecolor{darkred}{rgb}{0.5,0,0}
\pagecolor{white} % Background color
\color{black}     % Text color
\hypersetup{breaklinks=true, 
	colorlinks=true, 
	linkcolor=darkblue1, 
	menucolor=darkblue1, 
	urlcolor=darkblue1,
	citecolor=darkblue1,
	pdftitle={},
	pdfauthor={},
	pdfsubject={},
	pdfkeywords={},
	pdfproducer={}
}
\usepackage{siunitx}

\bibstyle{plain}

\begin{document}
	
%	\twocolumn[
	{%
		\begin{@twocolumnfalse}
			
			\begin{flushright}
				\normalsize
				%      \today
			\end{flushright}
			
			\vspace{-2cm}
			\title{\Large\boldmath Modeling Uncertainties on the Z Boson Background in the Context of High Precision W Boson Mass Measurements}

\author{Maarten Boonekamp$^{1}$, Matthias Schott$^{2}$, Chen Wang$^{3}$}
\institute{
\inst{1} l'Université Paris-Saclay, le CEA Paris-Saclay, France\\
\inst{2} Rheinische Friedrich-Wilhelms-University, Bonn, Germany\\ 
\inst{3} Institute of Physics, Johannes Gutenberg University, Mainz, Germany
}
% \noaffiliation

			\abstract{
The precise determination of the $W$ boson mass, $m_W$, is a cornerstone of electroweak precision tests and a critical input to global fits of the Standard Model (SM). While most existing measurements of $m_W$ are in agreement with SM predictions, the recent result from the CDF collaboration exhibits a significant deviation from the SM prediction but also exceeding $4\sigma$ from the global average of all other high precision measurements. This prompts renewed scrutiny of potential sources of systematic bias.  In this study, we investigate the modeling of the $Z$ boson background, which is particularly relevant in the muon decay channel of the $W$ boson, where undetected muons can mimic missing transverse momentum. Using updated theoretical predictions and high-precision simulations, we construct reweighted templates to evaluate the impact of possible mismodeling in the $Z$ background. Our analysis suggests that such effects can induce a shift in the extracted $m_W$ value of up to 8~MeV in the muon decay channel. While this shift is non-negligible and exceeds the originally quoted uncertainty from this source, it remains insufficient to explain the full discrepancy observed in the CDF measurement.}

	\maketitle
	\end{@twocolumnfalse}
}
%]

\tableofcontents

%%%%%%%%%%%%%%%%%%%%%%%%%%%%%%%%%%%%%%%%%%%%%%%%%%%%%%%%%%%%%%%%%%%%
\section{Introduction}

For decades, the precise measurement of the $W$ boson mass, $m_W$, has played a crucial role in testing the internal consistency of the Standard Model (SM) and in constraining potential contributions from new physics through global electroweak fits~\cite{gfitter}. The global electroweak fit combines measurements of key electroweak observables—particularly the masses of the $Z$ boson, the Higgs boson, and the top quark, as well as the effective weak mixing angle—to predict $m_W$ with remarkable precision. A precise and accurate measurement of $m_W$ thus provides a stringent test of the SM and can potentially reveal signs of physics beyond the Standard Model (BSM).

Most recent $m_W$ measurements are broadly consistent with each other and align well with the predictions of the global electroweak fit. However, the latest measurement from the CDF collaboration at the Tevatron deviates significantly, showing a discrepancy of over 5 standard deviations from the SM prediction but also a 4 sigma discrepancy to the average value of all other high precision measurements. This anomaly has sparked considerable interest in the high-energy physics community and has led to renewed scrutiny of both theoretical interpretations and the detailed methodology underlying $m_W$ measurements, mostly regarding its interpretation and the phenomenological aspects of the measurement procedure. The second category notably includes investigations into perturbative QCD corrections~\cite{Isaacson:2022rts}, polarization and parton distribution function (PDF) effects~\cite{LHC-TeVMWWorkingGroup:2023zkn}, and double parton scattering contributions~\cite{Zhang:2024hzr}. Each of these effects has been shown to potentially impact the extracted $m_W$ value by ${\cal O}(10~\mathrm{MeV})$. These discussions emphasize the need for a comprehensive re-evaluation of all systematic inputs contributing to precision $m_W$ measurements.

In this paper, we focus on a source of uncertainty that has received relatively less attention: the modeling of the $Z$ boson background. While typically assigned a small uncertainty in $m_W$ measurements, this background can be non-negligible especially in the muon decay channel. Events with undetected muons can mimic missing transverse energy and thus contaminate the $W \rightarrow \mu\nu$ signal region. Given the relatively hard kinematic spectrum of $Z$ boson events, mismodeling in their rate or shape can introduce a bias in the extracted $m_W$ value, particularly in template-based fits. We aim to systematically evaluate the impact of $Z$ boson modeling uncertainties on the $m_W$ measurement, with a particular emphasis on their potential contribution to the discrepancy observed by CDF. To this end, we use updated theoretical predictions, generate high-statistics Monte Carlo templates, and reweight $Z$ boson background shapes to assess the resulting bias in pseudo-data experiments.

The structure of this paper is as follows: 
Section~\ref{sec:recentmw} provides an overview of recent $m_W$ measurements, highlighting both consistent results and the notable CDF anomaly, in order to give a brief general introduction on the measurement methodology. Section~\ref{sec:modeling} describes our signal and background modeling approach, including the simulation setup and systematic variations as well as a discussion of the impact of $Z$ background mismodeling on the extracted $m_W$ value. Finally, Section~\ref{sec:conclusion} summarizes our findings and discusses the implications for future high-precision electroweak measurements.

%%%%%%%%%%%%%%%%%%%%%%%%%%%%%%%%%%%%%%%%%%%%%%%%%%%%%%%%%%%%%%%%%%%%

\section{Brief Review of high precision W Boson Mass Measurements \label{sec:recentmw}}

The measurement of \mW{} at hadron colliders has been a central pursuit in particle physics, with significant progress made over the past decades. Until 2007, the world average of \mW{} was largely dominated by measurements from the Large Electron–Positron (LEP) Collider. This changed with the CDF Collaboration at the Tevatron, which produced the first precision measurement of \mW{} with a comparable level of accuracy \cite{CDF:2012gpf}. Since then, the CDF and D0 Collaborations at the Tevatron \cite{D0:2012kms, CDF2TeV, CDF:2013dpa}, along with the ATLAS \cite{ATLAS7TeV2}, CMS \cite{CMS13TeV}, and LHCb \cite{LHCb13TeV} Collaborations at the Large Hadron Collider (LHC), have significantly advanced our understanding of \mW{} through increasingly precise measurements.

A major challenge in \mW{} measurements at hadron colliders is the incomplete reconstruction of the W boson decay kinematics. Unlike in electron–positron colliders, the initial parton collision energies in hadron collisions are unknown, limiting the reconstruction to the transverse plane relative to the beam axis. As a result, measurements focus on the electron and muon decay channels ($W \rightarrow e\nu$ and $W \rightarrow \mu\nu$) to avoid the significant multi-jet backgrounds present in hadronic decay channels.

The three main observables sensitive to \mW{} are: (i) the transverse momentum of the charged decay lepton, \pTl{}; (ii) the missing transverse energy, corresponding to the transverse momentum of the decay neutrino, \pTnu{}; and (iii) the transverse mass distribution, \mT{}. Among these, \mT{} is particularly robust against modeling uncertainties but is limited by the experimental resolution of the hadronic recoil-especially in high pile-up environments at the LHC. Extracting \mW{} involves a template fitting procedure in which theoretical predictions for the relevant observables are generated for different assumed values of \mW{}. The measured mass is determined by comparing these predictions to experimental data, typically using a $\chi^2$ minimization method or a global profile likelihood (PLH) fit. The PLH fit offers the advantage of simultaneously optimizing \mW{} along with nuisance parameters, thereby constraining systematic uncertainties using the data itself. Precision measurements of \mW{} require accurate modeling of both the detector response and the production and decay dynamics of the $W$ boson. Addressing these challenges is essential for reducing uncertainties and achieving a precise determination of \mW{} at hadron colliders.

This study investigates the $Z$ boson background, which can mimic $W^\pm \rightarrow \ell^\pm \nu$ decays in events where one lepton from the $Z$ decay escapes detection. Such events can exhibit apparent missing transverse momentum, resembling the signature of a genuine $W$ boson decay. At both the Tevatron and the LHC, the extensive coverage of electromagnetic and hadronic calorimeters ensures that most electrons from $Z$ decays are detected, significantly reducing the likelihood of fake missing transverse momentum in the electron channel. However, the situation is different for muons. Due to the more limited coverage of the muon detection systems, especially in the forward regions, muons have a higher probability of escaping detection. As a result, the $Z$ boson background contributes more significantly to the muon channel than to the electron channel. For instance, in the most recent CDF measurement, the $Z$ boson background accounts for only 0.134\% in the electron channel, compared to 7.37\% in the muon channel.

\begin{figure}
    \centering
	\subfigure[]{\includegraphics[width = 0.32\textwidth]{ 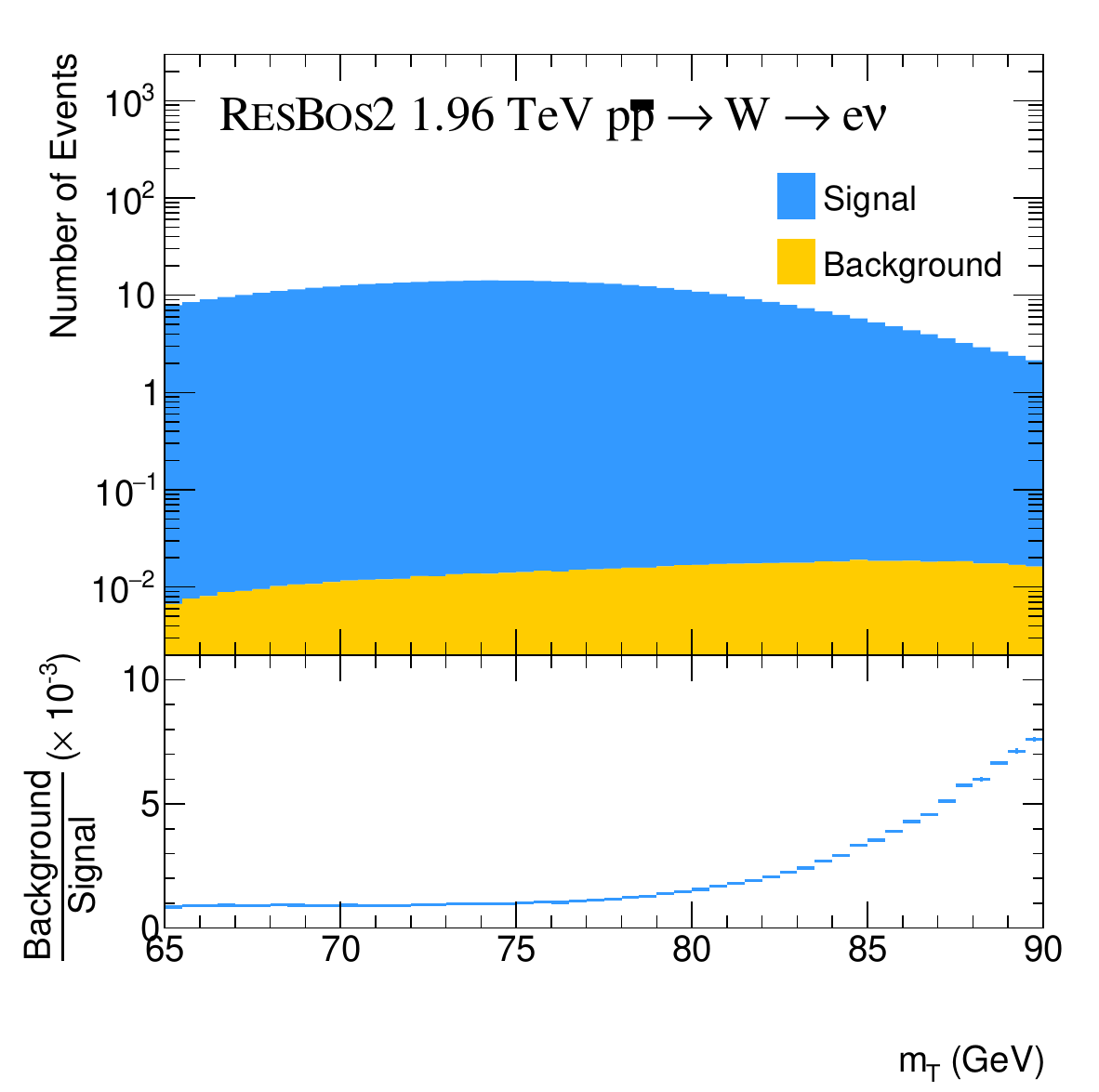}}
	\subfigure[]{\includegraphics[width = 0.32\textwidth]{ 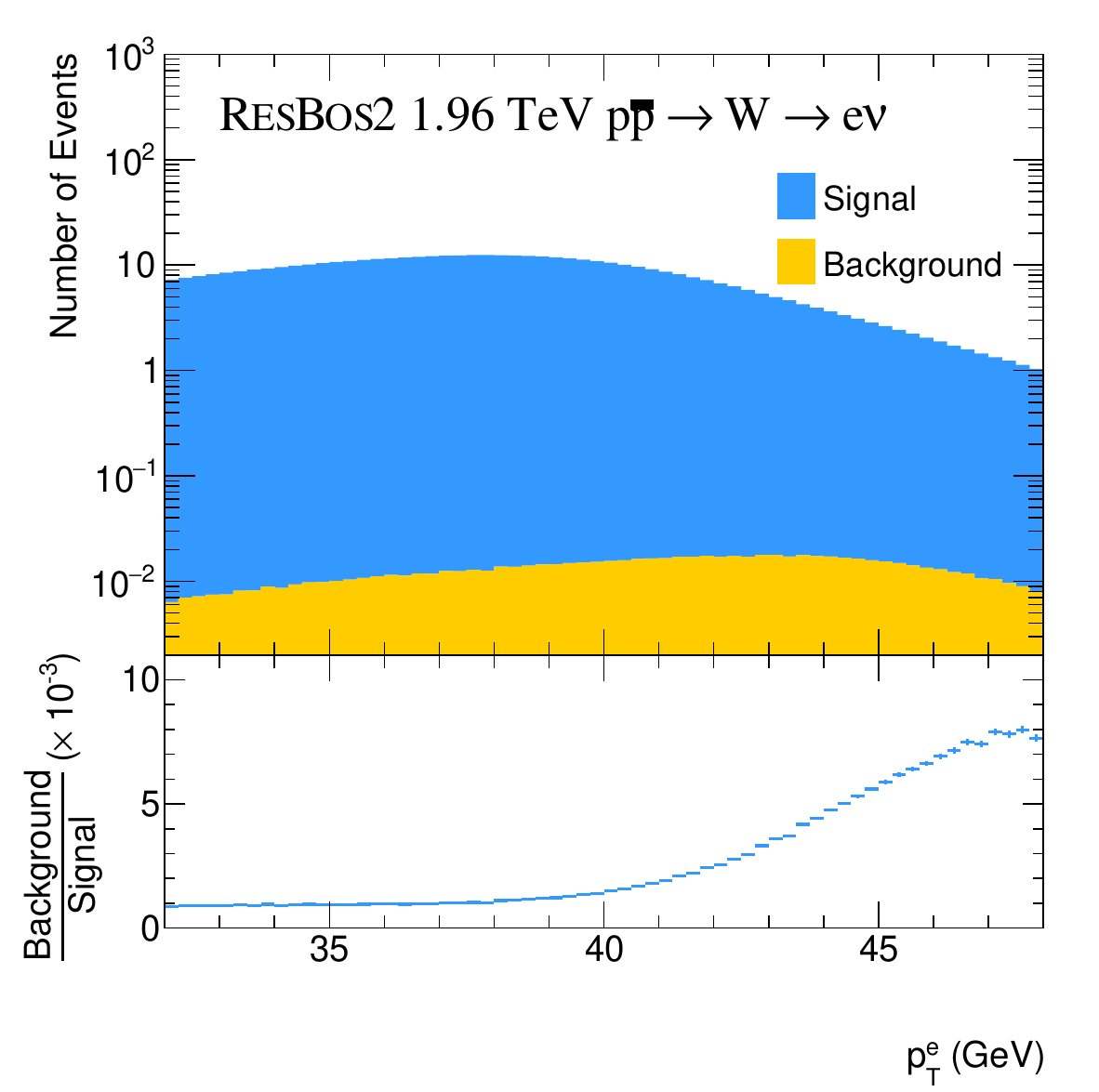}}
	\subfigure[]{\includegraphics[width = 0.32\textwidth]{ 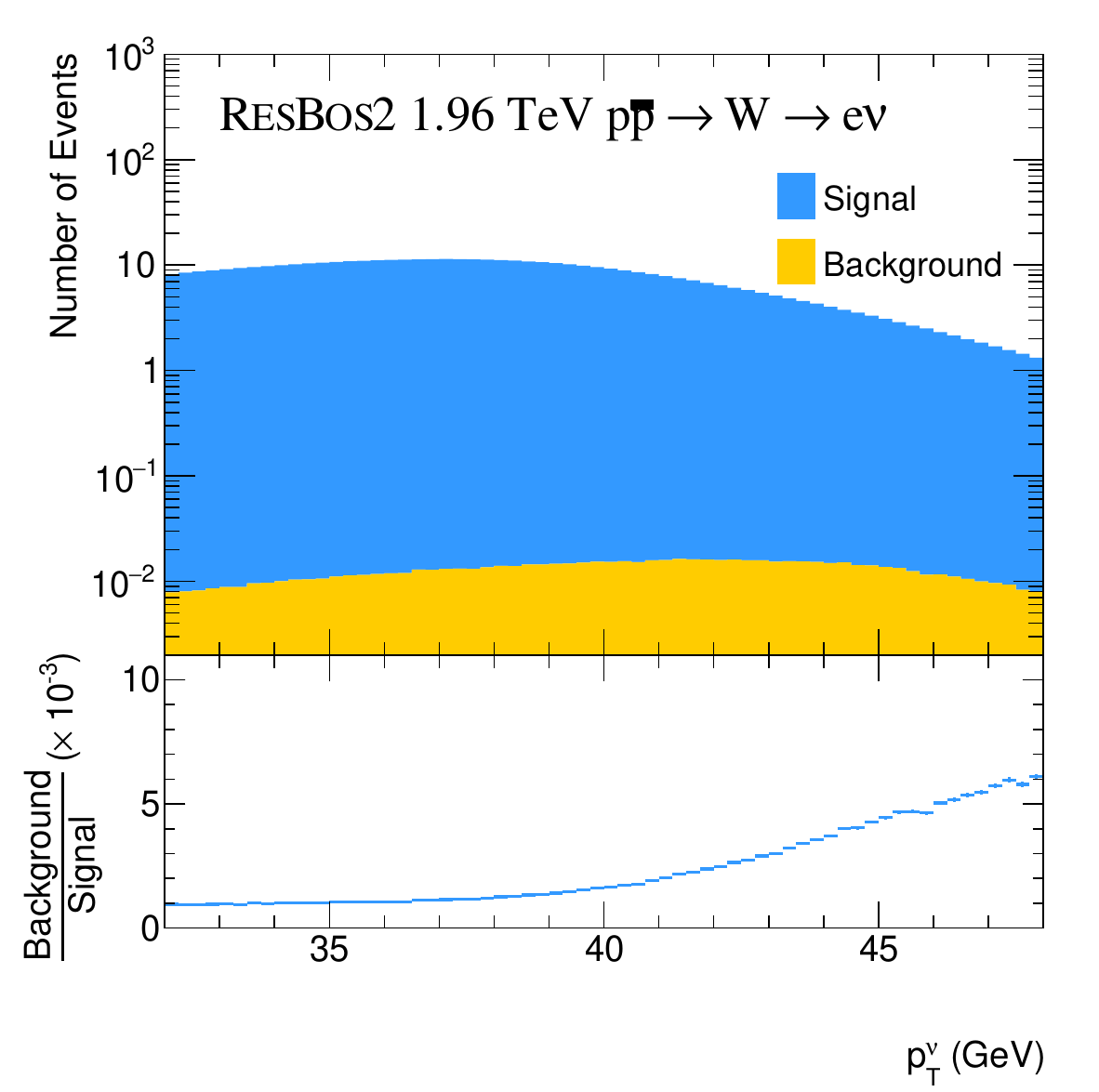}}\\
	\subfigure[]{\includegraphics[width = 0.32\textwidth]{ 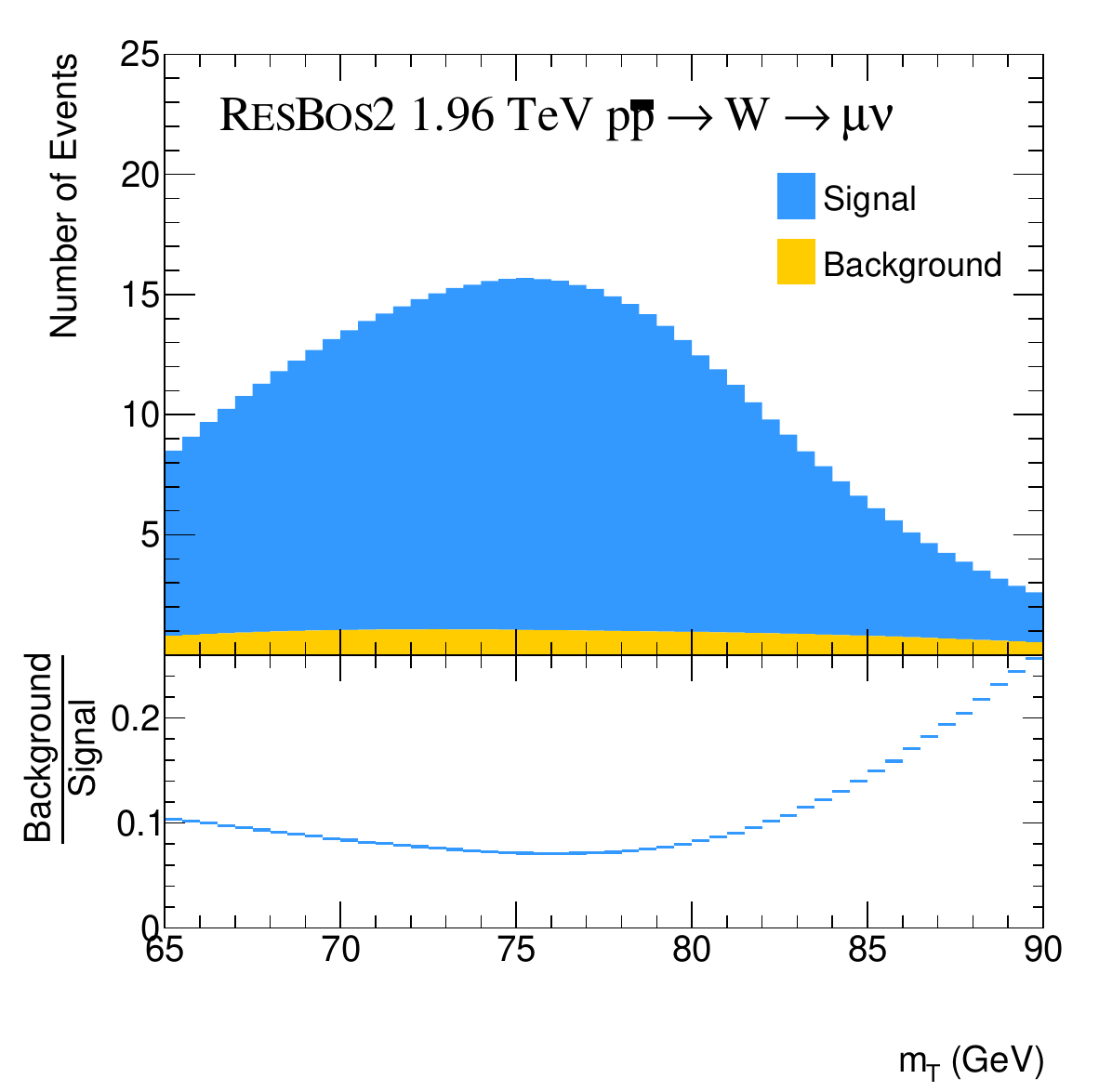}}
	\subfigure[]{\includegraphics[width = 0.32\textwidth]{ 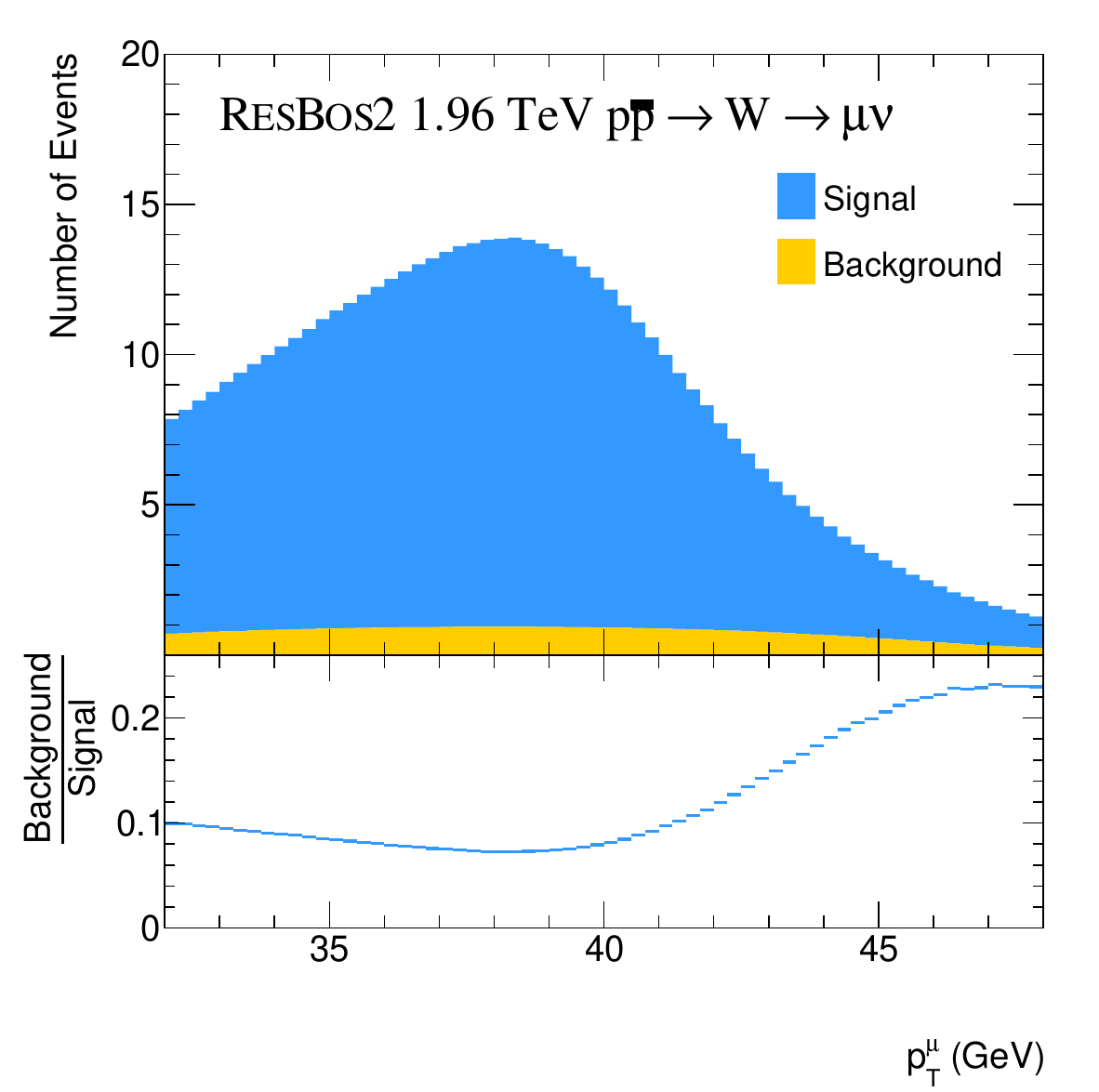}}
	\subfigure[]{\includegraphics[width = 0.32\textwidth]{ 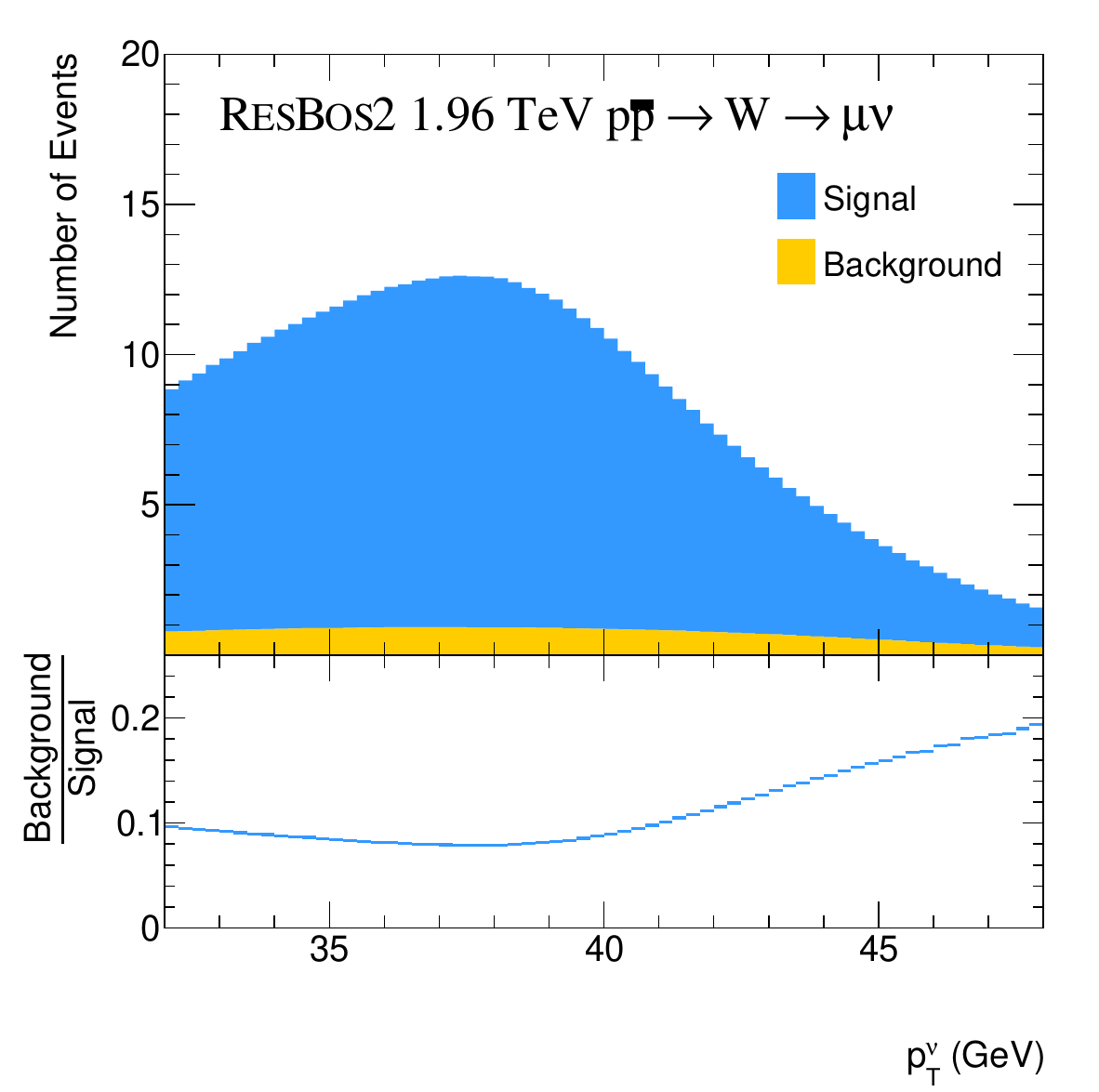}}
    \caption{The signal $W\rightarrow l\nu$ distribiutions (red) and the background $Z\rightarrow \ell\ell$ distributions (Blue) in the electron channel (top) and the muon channel (bottom) are show on the top panel and the corresponding background fraction is shown on the bottom panel. (a/d) are the \mT{} distributions, (b/e) are the \pTl{} distributions and (c/f) are the \pTnu{} distributions.}
\label{fig:sign_vs_bkg}
\end{figure}

\begin{figure}
        \centering
	\subfigure[]{\includegraphics[width = 0.40\textwidth]{ 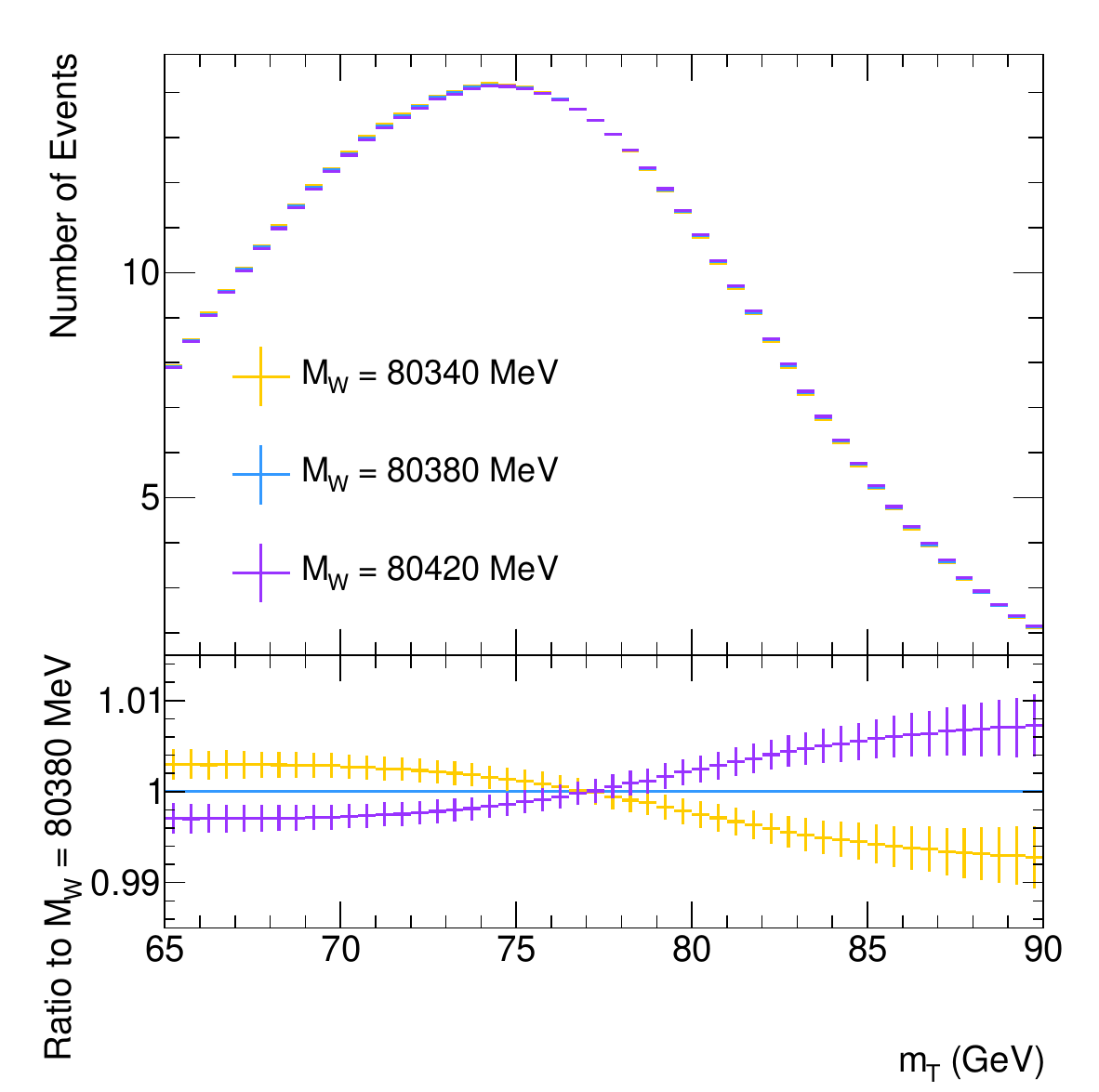}}
        \subfigure[]{\includegraphics[width = 0.40\textwidth]{ 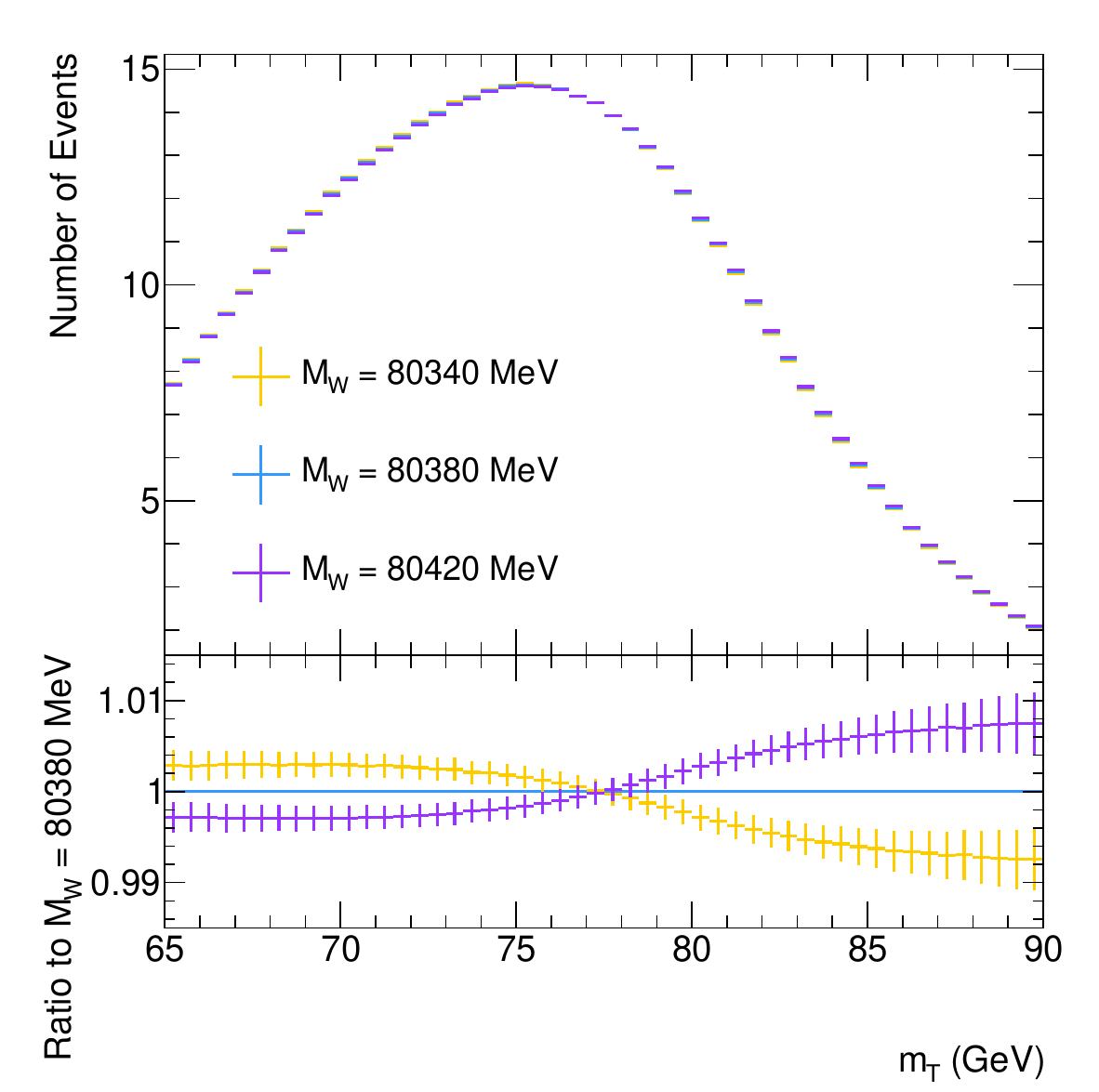}} \\
	\subfigure[]{\includegraphics[width = 0.40\textwidth]{ 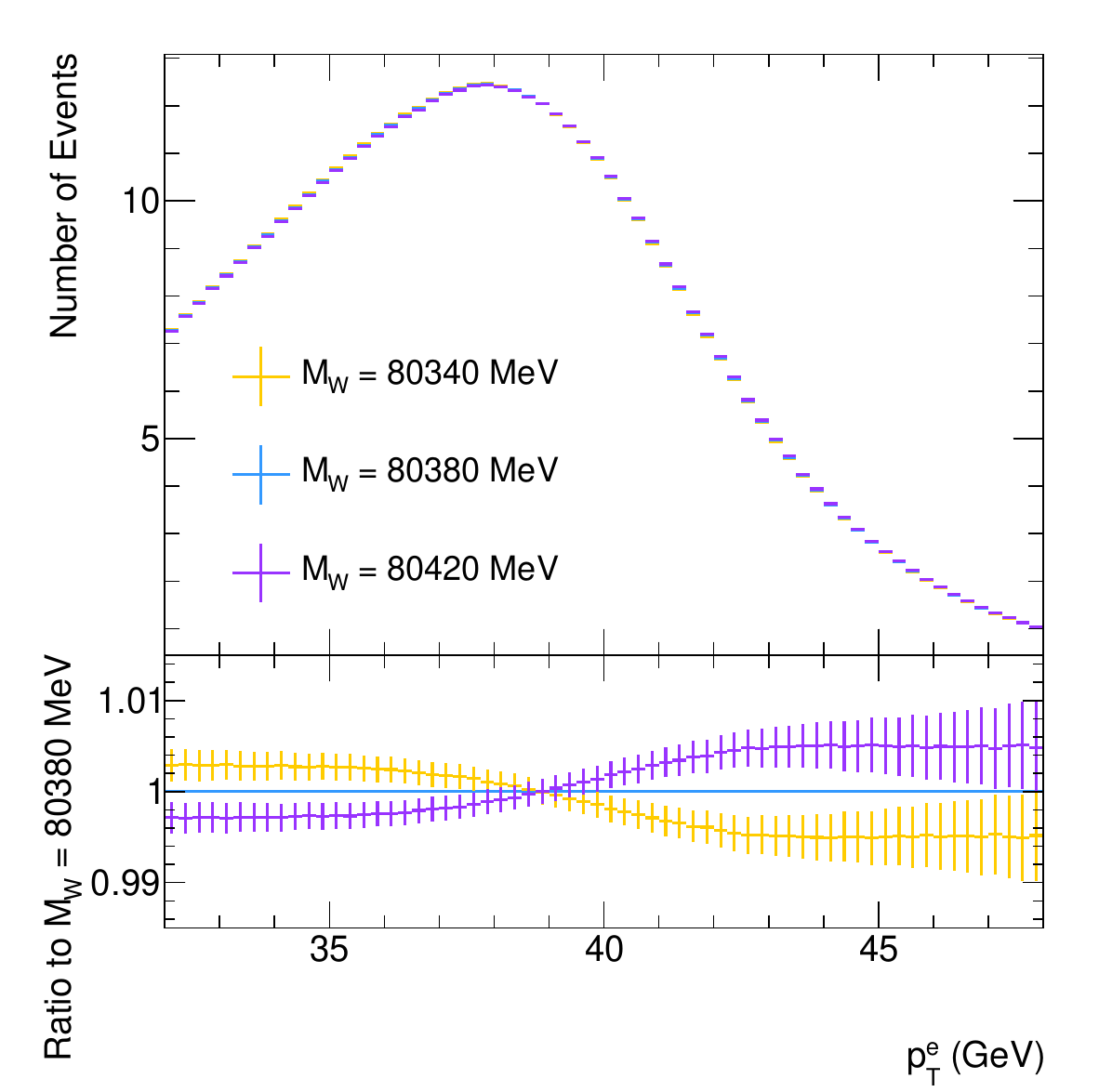}}
        \subfigure[]{\includegraphics[width = 0.40\textwidth]{ 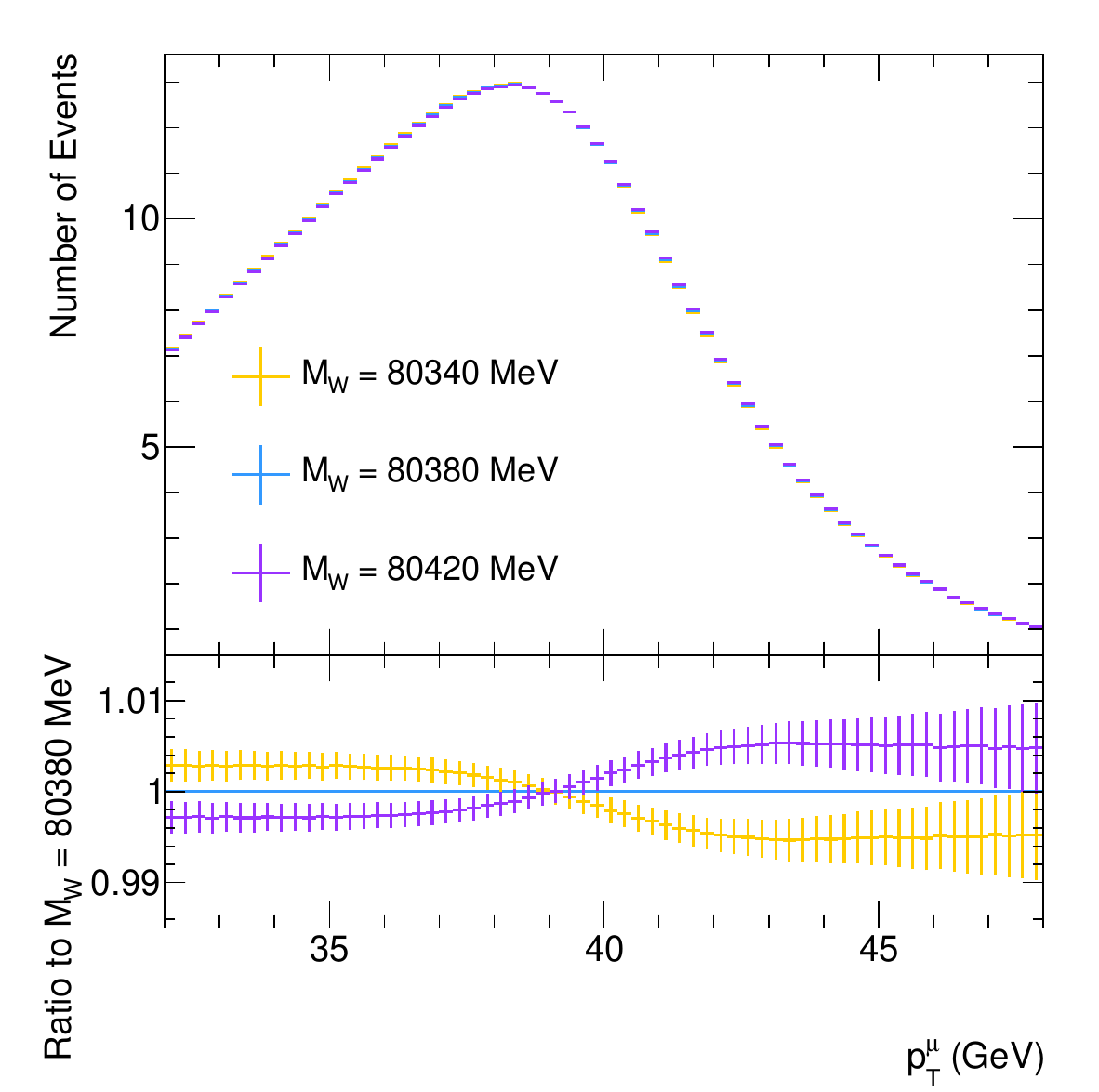}} \\
	\subfigure[]{\includegraphics[width = 0.40\textwidth]{ 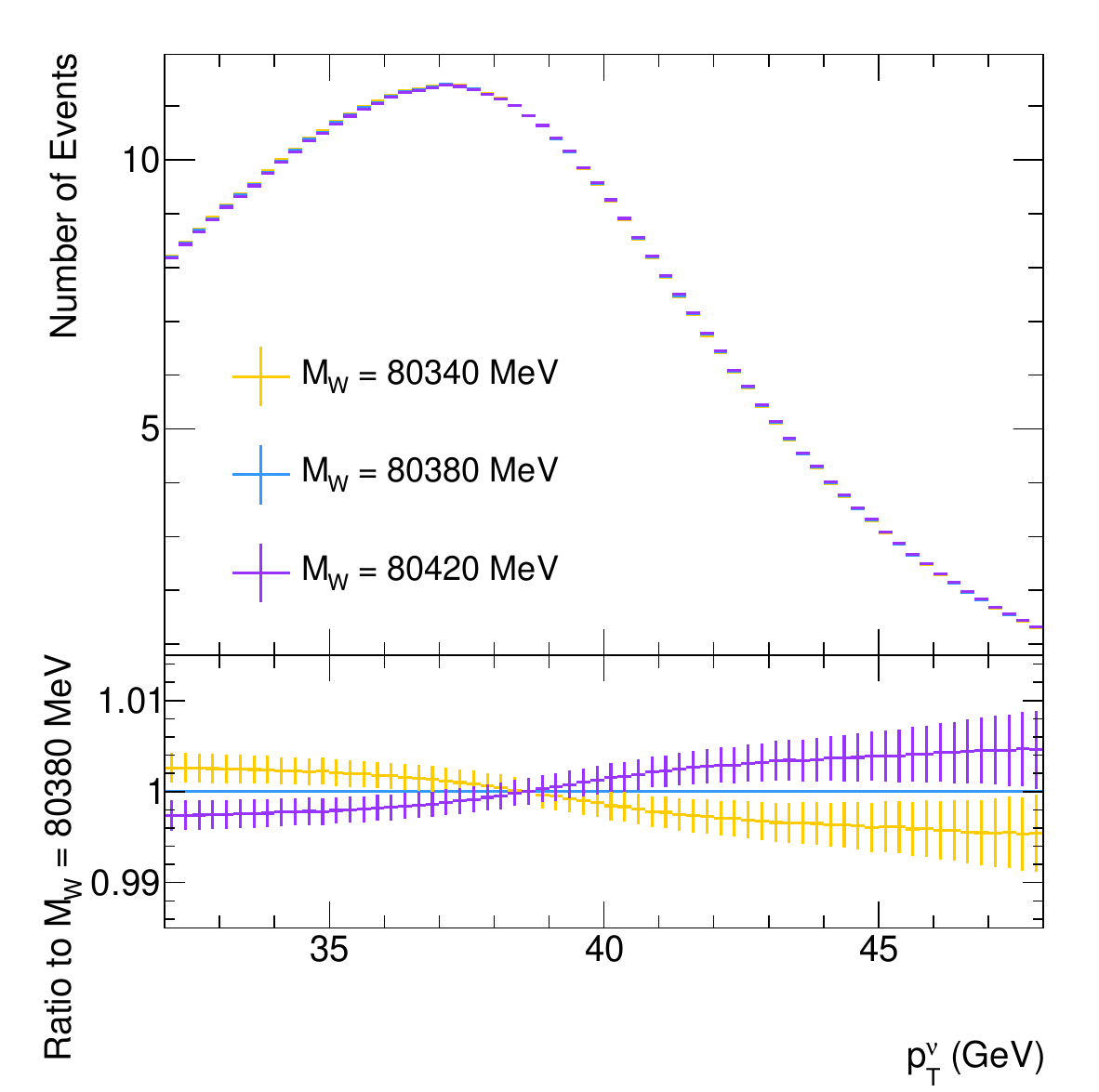}}
        \subfigure[]{\includegraphics[width = 0.40\textwidth]{ 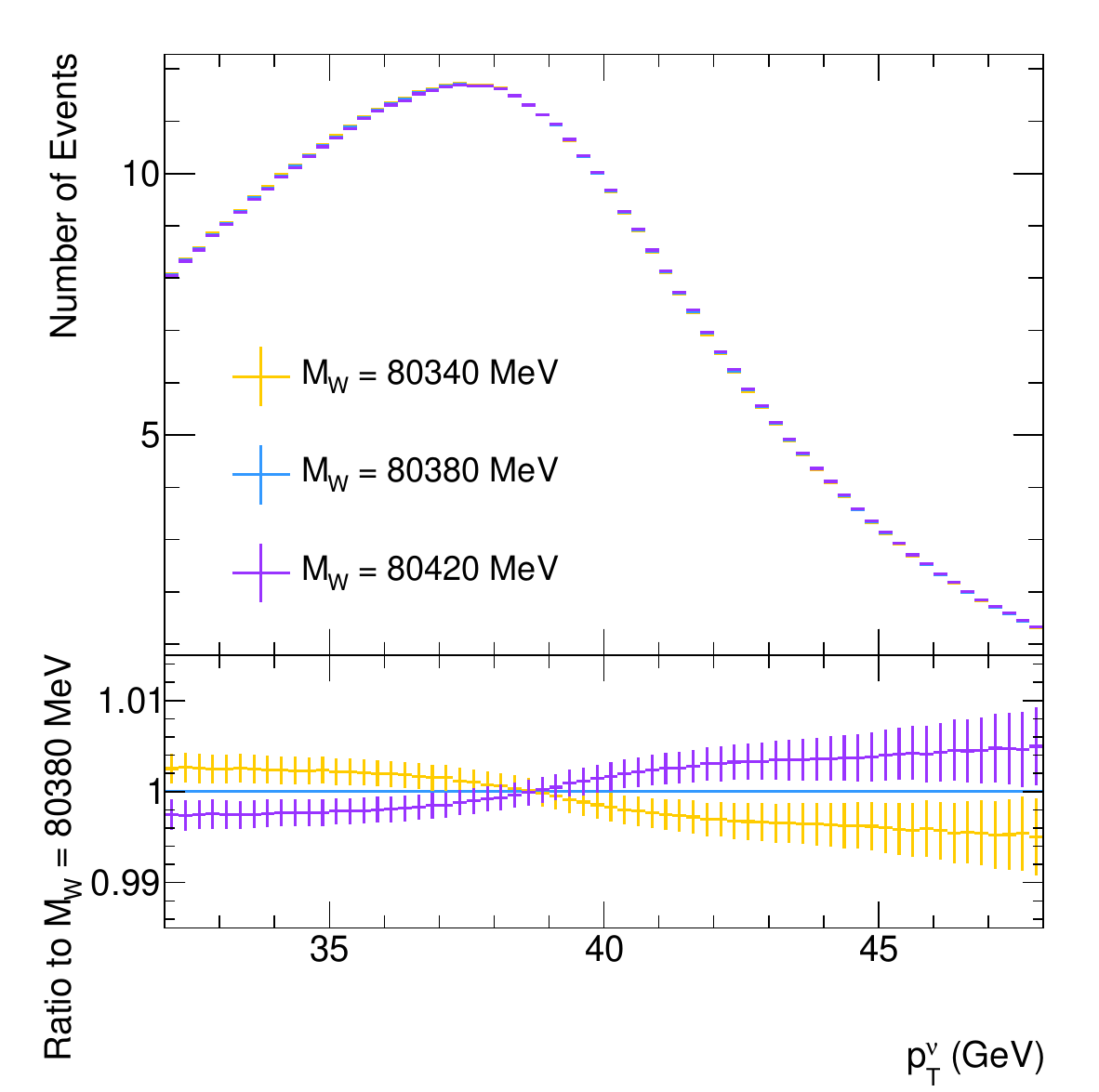}}
	\label{fig:sign_vs_mw}
	\caption{The \mT{} (a/b), \pTl{} (c/d) and \pTnu{} (e/f) distributions from the simulated signal samples reweighted to different $M_W$ values. The left three distributions are for the electron channel and the right three distributions are for the muon channel.}
\end{figure}

The contribution of the $Z$ boson background to the $W$ boson signal at the approximate reconstruction level of the CDF detector—modeled using a fast simulation developed by the LHC-Tevatron $M_W$ Working Group \cite{LHC-TeVMWWorkingGroup:2023zkn}—is illustrated in Fig.~\ref{fig:sign_vs_bkg}. The figure shows the distributions of the lepton transverse momentum (\pTl{}), the neutrino transverse momentum (\pTnu{}), and the transverse mass (\mT{}) for both the electron and muon decay channels in $p\bar{p}$ collisions at a center-of-mass energy of 1.96 TeV. Fig.~\ref{fig:sign_vs_mw} further illustrates the effect of varying the $W$ boson mass (\mW{}) by $\pm10$~MeV and $\pm40$~MeV, highlighting the sensitivity of the measurement to the $Z$ boson background.

In the following, we provide an overview of the measurement strategies employed by the ATLAS, LHCb, CMS, and CDF collaborations to achieve the most precise determinations of the $W$ boson mass, \mW{}. The individual results, along with the breakdowns of statistical and systematic uncertainties, are summarized in Table~\ref{tab:mw_results}.

\begin{table}[ht]                             
\centering
\begin{tabular}{l|rr|r|rrrrr|r}
\hline
Experiment & \multicolumn{2}{c|}{Collision}& $\mathcal{L}$(fb$^{-1}$)  & $m_W$ (MeV) & Stat. Unc. & Syst. Unc. & Bkg. Unc.& Total Unc. & Pull\\
\hline
\multirow{2}{*}{D0 \cite{D0:2012kms} }& \multirow{2}{*}{$p\bar{p}$} & \multirow{2}{*}{1.96~TeV} & 4.3 & 80367 & 13 & 22 &2& 26 & 0.4\\
& & & 5.3 & 80375 & 11 & 20 & - &23 & 0.8\\
\hline
LHCb \cite{LHCb13TeV} & $pp$ & 13~TeV & 1.7 & 80354 & 23 & 22 & $<1$ & 32 & -0.1\\ 
\hline
\multirow{2}{*}{CDF \cite{CDF:2012gpf, CDF2TeV}} & \multirow{2}{*}{$p\bar{p}$} & \multirow{2}{*}{1.96~TeV} & 2.2 & 80387 & 12 & 15 & 3 & 19 & 1.5\\
& & & 8.8 & 80433.5 & 6.4 & 6.9 & 3.3& 9.4 & 6.3 \\
\hline
ATLAS \cite{ATLAS7TeV2} & $pp$ & 7~TeV & 4.6 & 80366.5 & 9.8 & 12.5 & 2.0 & 15.9 & 0.6\\
\hline
CMS \cite{CMS13TeV} & $pp$ & 13~TeV & 16.8 & 80360.2 & 2.4 & 9.6 & - & 9.9 & 0.3\\
\hline
\end{tabular}
\caption{Measurements of $m_W$ from different experiments, D0, CDF, LHCb, ATLAS and CMS, The last column contains the pull comparing to the latest indirect determination from {\sc GFitter}}
\label{tab:mw_results}
\end{table}

\subsection{ATLAS}

The ATLAS experiment has performed two high-precision measurements of the $W$ boson mass, \mW{}, using proton-proton collision data collected in 2011 at a center-of-mass energy of 7 TeV, corresponding to an integrated luminosity of 4.6 fb$^{-1}$. Both measurements determine \mW{} using the transverse momentum of the decay lepton, \pTl{}, and the transverse mass, \mT{}, distributions in the electron and muon decay channels. The analysis includes leptons within a pseudorapidity range of $|\eta| < 2.4$, with energy and momentum calibrations performed using $Z$ boson events.

The first ATLAS measurement \cite{ATLAS7TeV} employed a simple $\chi^2$ minimization technique, resulting in a measured value of \mW{} = $80370 \pm 19$ MeV. This approach analyzed the \pTl{} and \mT{} distributions separately for positively and negatively charged leptons, dividing the data into three to four distinct $\eta$ regions. This strategy helped reduce systematic uncertainties and enabled consistency checks across different event categories.

An updated measurement \cite{ATLAS7TeV2} utilized the same dataset and calibrations but introduced a profile likelihood (PLH) fit method and updated parton distribution function (PDF) sets. The new result, $m_W = 80360 \pm 16$ MeV, demonstrated an approximate 20\% reduction in total uncertainty. The PLH fit more effectively constrains several sources of systematic uncertainty compared to the $\chi^2$ method.

In both analyses, $W$ and $Z$ boson templates were generated using the {\sc Powheg} program \cite{Powheg,Powheg2,Powheg3} with the {\sc CT10} PDF set \cite{CT10}, and full detector simulation was performed using the \textsc{Geant4} framework \cite{Geant4}. These templates were further refined through reweighting techniques to account for accurate modeling of angular coefficients $A_i$ at next-to-next-to-leading order (NNLO) precision, and to incorporate variations across different PDF sets using predictions from the \textsc{DYTurbo} program \cite{DYTurbo}. Reweighting also improved the modeling of the transverse momentum distributions of the $W$ and $Z$ bosons, with the baseline predictions tuned to the measured $Z$ boson spectrum. Detailed information on the template generation and reweighting procedure can be found in Ref.~\cite{ATLAS7TeV2}.

An overview of the breakdown of ATLAS’s experimental uncertainties is provided in Table~\ref{tab:WUncer}. The uncertainty arising from background contributions adds 2.0 MeV to the total uncertainty on the $m_W$ measurement.

\begin{table}[h!]
\begin{tabular}{p{7cm}|cc|c|cc}
\hline
Experiment				&	\multicolumn{2}{c|}{ATLAS \cite{ATLAS:2023fsi}}	&	CMS \cite{LHCb:2021bjt} & \multicolumn{2}{c}{CDF \cite{CDF:2022hxs}}	\\
\hline
Observable				&	\multicolumn{1}{c|}{$p_T^{l}$[MeV]}	&	\multicolumn{1}{c|}{$m_T$ [MeV]}	& $p_T^{l}$[MeV] & \multicolumn{1}{c|}{$p_T^{l}$[MeV]}	&	\multicolumn{1}{c}{$m_T$ [MeV]}	\\
\hline
$m_W$					&	\multicolumn{1}{c|}{80360}	&	\multicolumn{1}{c|}{80382	}&	80360	& 80421 & 80439	\\
\hline
Stat. Unc.					 &	5	&	7	&	2.4	& 8	&	8	\\
Sys. Unc. 					 &	10	&	16	&	6.9	& 6	&	5	\\
Model Unc. 				     &	11	&	20	&	6.7 & 6	&	5		\\
\hline
Total Unc. 				     &	16	&	25	&	9.9 & 12 & 10 \\
\hline
Lepton Calib. Unc.			 &	8	&	9	&	5.6 & 2	&	2	\\
Had. Calib. Unc.			 &	1	&	12	&   0.0 & 4	&	2	\\
Z Boson background     		 &	X	&	X	&	X   & X &	X	\\
Other Exp. Unc.			     &	3	&	6	&	3.2 & 5	&	3	\\
PDF						     &	8	&	15	&	4.4 & 4	&	4	\\
EW\&QED Effects			     &	6	&	6	&   2.0 & 3	&	3	\\
$p_T(W)$ modelling			 &	5	&	10	&	2.0 & 2	&	1	\\
\hline
Final Result of Collaboration	&	\multicolumn{2}{c|}{$80360\pm16$}		&	$80360\pm10$	& \multicolumn{2}{c}{$80433\pm10$}		\\
(Stat., Exp. Sys., Model Unc.)	&	\multicolumn{2}{c|}{$(\pm5\pm10\pm11)$}	&	$(\pm2.4\pm6.9\pm6.7)$	& \multicolumn{2}{c}{$(\pm2.4\pm6.9\pm6.7)$}		\\
\hline
\end{tabular}
\centering
\caption{\label{tab:WUncer}Breakdown of Uncertainties for the $W$ boson mass measurements at the ATLAS, CMS and CDF Experiments. The Z boson background does not play a significant role for the measurements from the LHCb and the D0 Collaboration.}
\end{table}

\begin{table}[h!]
\scriptsize
\centering
\begin{tabular}{l|rrr|l|rrr|l|rrr}
			\hline
			PDF & $\sigma_W$ & $\sigma_Z$ & $R_{W/Z}$ & PDF & $\sigma_W$ & $\sigma_Z$ & $R_{W/Z}$ & PDF & $\sigma_W$ & $\sigma_Z$ & $R_{W/Z}$\\
			\hline
			MSTW2008 nnlo & 1375 & 250.9 & 10.95     & NNPDF2.3 nnlo & 1372 & 251.1 & 10.93 & 			CTEQ6.1 & 1315 & 243.09 & 10.82\\
			MMHT2014 nnlo & 1395 & 256.5 & 10.87     & NNPDF3.0 nnlo & 1350 & 247.4 & 10.91 & CT10 nnlo & 1369.05 & 252.9 & 10.83 \\
			MSHT20 an3lo & 1379 & 256.0 & 10.77      & NNPDF3.1 nnlo & 1392 & 256.9 & 10.84 & CT14 nnlo & 1378.68 & 254.1 & 10.85 \\
			MSHT20qed an3lo & 1372 & 254.8 & 10.77   & NNPDF4.0 nnlo & 1398 & 258.3 & 10.83 & CT18 nnlo & 1389.37 & 255.5 & 10.88\\
			\hline
		\end{tabular}
		\caption{Total cross sections of $p\bar{p}\rightarrow W^+ \rightarrow \mu^+ \nu$ and  $p\bar{p}\rightarrow Z/\gamma^*\rightarrow \mu^+ \mu^-$ predicted from N$^3$LO+N$^3$LL calculations by {\sc DYTurbo} with different PDF sets: MSTW2008 \cite{Martin:2009iq}, MMHT2014 \cite{Harland-Lang:2014zoa}, MSHT20 \cite{Bailey:2020ooq}, NNPDF2.3 \cite{Ball:2012cx}, NNPDF3.0 \cite{NNPDF:2017mvq}, NNPDF3.1 \cite{NNPDF:2017mvq}, NNPDF4.0 \cite{NNPDF:2021njg}, CTEQ6.1 \cite{Stump:2003yu}, CT10 \cite{Lai:2010vv}, CT14 \cite{Dulat:2015mca}, CT18 \cite{Hou:2019efy}}
		\label{tab:rwz_pdf}
\end{table}

\subsection{CMS}
The first measurement of the $W$ boson mass, \mW{}, by the CMS collaboration was performed using data collected in 2016 from proton-proton collisions at the LHC with a center-of-mass energy of $\sqrt{s} = 13$ TeV, corresponding to an integrated luminosity of 16.8 fb$^{-1}$\cite{CMS13TeV}. Simulated $W$ and $Z$ boson event samples were generated using the {\sc MiNNLO} framework\cite{Minnlo, Minnlo2}, with \textsc{Pythia8}\cite{Sjostrand:2014zea} for parton showering and hadronization, and \textsc{Photos++}\cite{Photos,Photos2} for final-state photon radiation, all based on the CT18Z PDF set~\cite{CT18Z}. The simulation achieves next-to-next-to-leading order (NNLO) plus next-to-next-to-next-to-leading-logarithmic (N3LL) accuracy for the transverse momentum distributions of the leptons, \pTl{}.

A key aspect of the CMS measurement is its momentum calibration strategy, which primarily relies on $J/\psi$ events. $Z$ boson events are used to assess uncertainties on the momentum scale and to apply efficiency corrections. Unlike the ATLAS and CDF measurements, the CMS analysis focuses exclusively on the muon decay channel and uses only the transverse momentum distribution of the muons, \pTl{}, avoiding both the transverse mass observable, \mT{}, and the electron decay channel. The \mW{} extraction is performed using a profile likelihood (PLH) fit in a finely binned, multidimensional space of the lepton’s transverse momentum, pseudorapidity, and charge. This approach allows for the simultaneous optimization of the \mW{} value and the nuisance parameters associated with systematic uncertainties, thereby improving the overall precision of the measurement.

The final CMS results, along with a detailed breakdown of uncertainties, are also presented in Table~\ref{tab:WUncer}. While the modeling and experimental uncertainties in the CMS measurement are comparable to those of ATLAS, the CMS analysis benefits from a significantly larger data sample, potentially reducing the statistical component of the uncertainty. However, the reduced set of internal consistency checks—due to the exclusive use of the muon channel and the \pTl{} observable—may limit the ability to fully validate the measurement.

\subsection{LHCb Experiment}

The LHCb detector is optimized for precision $b$-quark physics and is therefore designed as a forward detector system. Its \mW{} measurement is based on an integrated luminosity of 1.7 fb$^{-1}$, recorded in 2016, and uses only the muon decay channel, which is primarily calibrated using $Z \rightarrow \mu\mu$ events~\cite{LHCb13TeV}. Given the detector’s forward geometry, muons are reconstructed in the region $1.7 < |\eta| < 5.0$, thereby probing a different Björken-$x$ regime in the proton PDFs compared to central detectors.

The \mW{} value is determined through a simultaneous fit of the lepton transverse momentum distribution, \pTl{}, from $W$ boson candidate events and the $\Phi^*$ distribution from $Z$ boson candidate events. The $\Phi^*$ distribution serves as a proxy for the transverse momentum distribution of the $Z$ boson. The fit simultaneously determines \mW{} along with several additional parameters, including the fractions of $W^+$ and $W^-$ events and parameters describing the transverse momentum distribution of the vector bosons. The resulting $W$ boson mass is found to be $80354 \pm 32$ MeV.

\subsection{D0 Experiment}

The D0 experiment is based on half of the available dataset from Tevatron Run II and employs only the electron decay channel~\cite{D0:2012kms}. The lepton energy and momentum calibration is performed using the $Z$ boson mass peak, and the extraction of \mW{} is based on the inclusive \pTl{}, \mT{}, and \pTnu{} distributions for leptons reconstructed in the central region of the detector, defined by a pseudorapidity requirement of $|\eta| < 1.05$. Currently, there are no plans to analyze the remaining portion of the dataset. The final D0 measurement of \mW{} yields a value of $80375 \pm 23$ MeV. Given that the $Z$ boson background in the electron decay channel plays only a minor role, we refrain from a detailed discussion of the analysis.

\subsection{CDF Experiment}

The CDF experiment uses the full dataset from Tevatron Run II, corresponding to an integrated luminosity of 8.8 fb$^{-1}$, and analyzes both the electron and muon decay channels~\cite{CDF2TeV}. The track momentum calibration is primarily based on $J/\psi \rightarrow \mu\mu$ events, while the energy calibration is performed using the $E/p$ ratio from $Z \rightarrow ee$ and $W \rightarrow e\nu$ events. Similar to the D0 experiment, a $\chi^2$ minimization method was used for the \mW{} determination, employing both decay channels and the inclusive \pTl{}, \mT{}, and \pTnu{} distributions for leptons with $|\eta| < 1.0$, yielding a measured value of $80435 \pm 9$ MeV.

Several crucial aspects of the CDF measurement should be noted, as outlined in Ref.\cite{LHC-TeVMWWorkingGroup:2023zkn}. In the first step, Monte Carlo simulations for both signal and background processes were performed using an early version of the \textsc{ResBos} program\cite{Resbos}, operating at next-to-leading order (NLO+NLO) accuracy and employing the CTEQ6M PDF set~\cite{CTEQ6M}. Following the generation of these samples, the CDF detector response was simulated. The resulting signal and background templates served as the basis for the initial $W$ boson mass determination, using a fitting procedure to match the templates to experimental data.

As far as we understand, in a second step the updated signal templates were generated using the second-generation \textsc{ResBos2} program~\cite{ResBos2}, achieving next-to-next-to-leading-logarithm and next-to-next-to-leading-order (NNLL+NNLO) accuracy. These newer templates were based on the more advanced {\sc NNPDF3.1} set \cite{NNPDF:2017mvq}, while the non-perturbative parameters governing the low-energy behavior of the vector boson transverse momentum spectrum were tuned at next-to-leading-logarithm (NLL) accuracy using the original {\sc CTEQ6M} set \cite{Nadolsky:2008zw}. These updated signal templates were used to correct the measurement in three aspects in one go, which were the update of PDF from {\sc CTEQ6M} to {\sc NNPDF3.1}, the correction in the polarization and the removal of a particle level selection on the mass of $W$ boson, $m_{\ell\nu}<150~$GeV. The sum of these effects is about 3-4 MeV as quoted by CDF Collaboration, but results from the compensation of around 10~MeV PDF and polarization shifts.

To quantify the differences between the original and updated templates, the CDF collaboration compared the transverse momentum of the lepton (\pTl{}), the transverse momentum of the neutrino (\pTnu{}), and the transverse mass (\mT{}) distributions from the original analysis with those predicted by the new signal templates. From this comparison, a shift was derived, and the corresponding uncertainties on the measured $W$ boson mass were calculated and applied to the final result. An exhaustive summary of both experimental and modeling uncertainties—including those related to background estimation—is provided in Table~\ref{tab:WUncer}.

%%%%%%%%%%%%%%%%%%%%%%%%%%%%%%%%%%%%%%%%%%%%%%%%%%%%%%%%%%%%%%%%%%%%

\section{Estimation of Modeling Uncertainties on the Z Boson Background \label{sec:modeling}}

\subsection{Scientific Objective}

There are several key aspects of the original CDF analysis that require additional scrutiny, particularly the $Z$ boson background templates, which might not have been updated in parallel with the signal templates. Specifically, the background modeling appears to rely on the first version of the \textsc{ResBos} program at NLL+NLO accuracy, using the {\sc CTEQ6M} PDF set. Furthermore, the estimated fraction of $Z$ boson events subtracted from the data is based on an assumed $W$-to-$Z$ boson production cross-section ratio derived from potentially outdated theoretical predictions. These factors introduce several potential issues that could affect the accuracy of the \mW{} measurement:

\begin{itemize}
	\item The initial version of \textsc{ResBos} predicts incorrect $A_i$ coefficients shown in Fig.~6 of Ref.~\cite{LHC-TeVMWWorkingGroup:2023zkn}, which may distort both the shape of the $Z$ boson background distribution and its acceptance by the detector. 
	\item The outdated {\sc CTEQ6M} PDF set affects not only the acceptance of the detector but also the expected shape of the $Z$ boson background distributions.
	\item The assumed number of $Z$ boson background events, or the $W$/$Z$ cross-section ratio, is based on an older prediction derived using an invariant mass range of the dilepton system between 66 and 116 GeV, potentially biasing the predicted $Z$ boson event rate.
\end{itemize}

The aim of this paper is to quantify the impact of these factors on the measured value of \mW{} reported by the CDF collaboration. It is important to note, however, that it remains uncertain whether the CDF analysis strictly followed the steps discussed here%, as the original publication lacks sufficient detail to confirm this. 
These potential issues are expected to predominantly affect the muon channel, where the contribution from the $Z$ boson background is more significant, while the electron channel is likely to remain largely unaffected.

\subsection{Scientific Methodology}

The effect of the $Z$ boson background on the $W$ boson mass measurement of the CDF collaboration is estimated by employing a detailed and systematic approach. First, we use a state-of-the-art N$^3$LO+N$^3$LL prediction of the $W$ and $Z$ boson cross-sections, applying a dilepton mass requirement from 66 to 116\,GeV for the $Z$ boson process and utilizing the {\sc MSTW2008} PDF set \cite{Martin:2009iq}. To accurately model the shapes of the $W$ and $Z$ boson templates, we use {\sc ResBos2} at NNLL+NNLO accuracy, with non-perturbative parameters taken from Ref.~\cite{RunITune}. This ensures a realistic description of the kinematic distributions, particularly in the low transverse momentum region, which is critical for the $W$ boson mass measurement. The detector response is then simulated using the framework from Ref.~\cite{LHC-TeVMWWorkingGroup:2023zkn}, allowing us to account for the acceptance and resolution effects of the CDF detector. With the signal and background templates prepared, we combine them according to the predicted cross-sections and acceptances to generate pseudo-data samples. This approach creates a controlled environment in which the effects of mismodeling the $Z$ boson background can be isolated and studied.

To assess the impact of the $Z$ boson background, we generate alternative $Z$ boson background templates using {\sc ResBos2} with different accuracies and various PDF sets. These new templates are subtracted from the pseudo-data, simulating the procedure used in a real analysis. The remaining pseudo-data is then fitted with the original $W$ boson template, allowing the $W$ boson mass to vary freely.

The difference between the nominal $W$ boson mass value used to generate the pseudo-data and the value extracted from the fit provides a quantitative estimate of the bias introduced by potential mismodeling of the $Z$ boson background. This method not only quantifies the size of the effect but also highlights the sensitivity of the measurement to different sources of systematic uncertainty, particularly those related to $Z$ boson modeling.

\begin{figure}
	\centering
	\subfigure[]{\includegraphics[width = 0.49\textwidth]{ 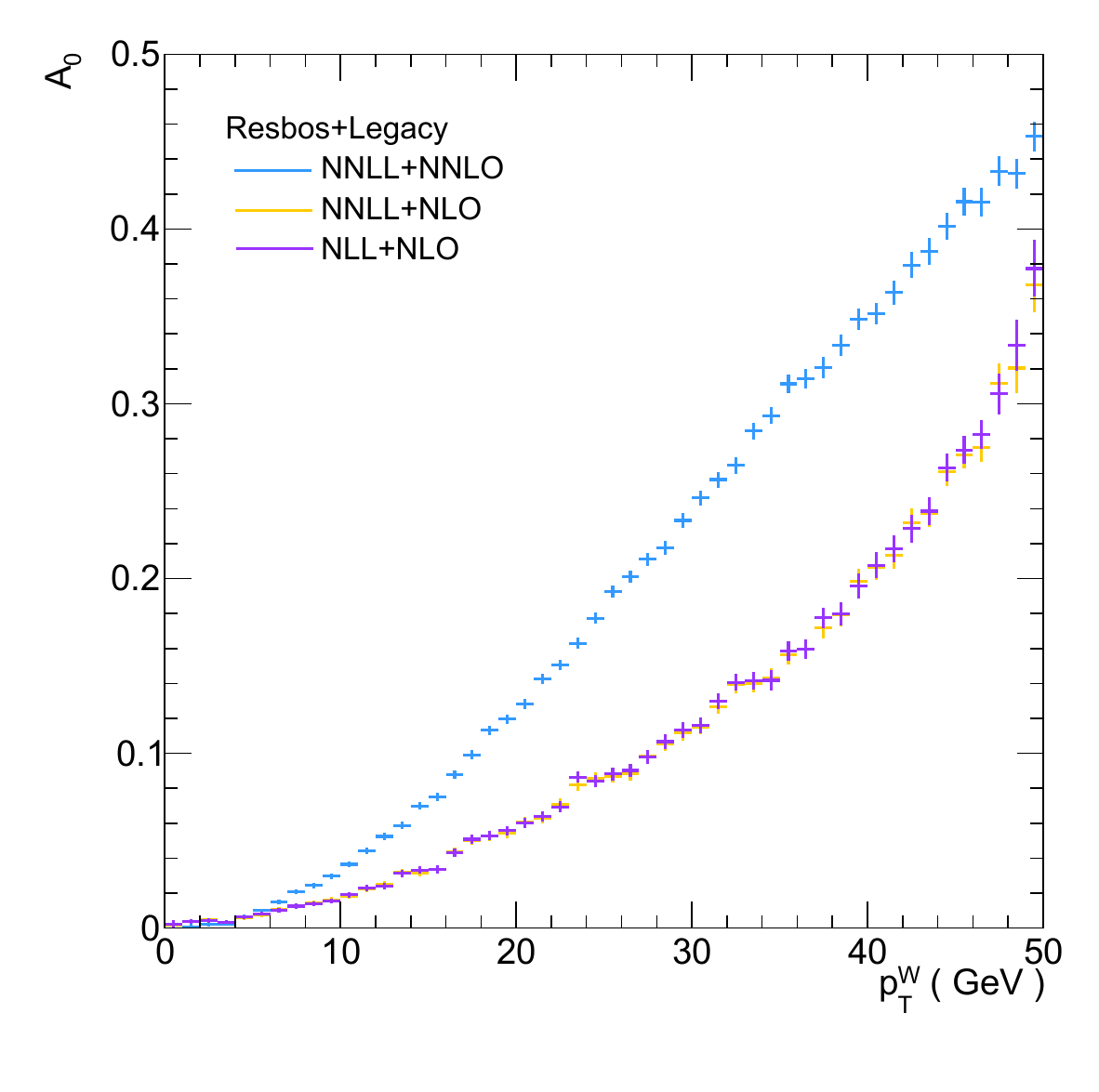}}
	\subfigure[]{\includegraphics[width = 0.49\textwidth]{ 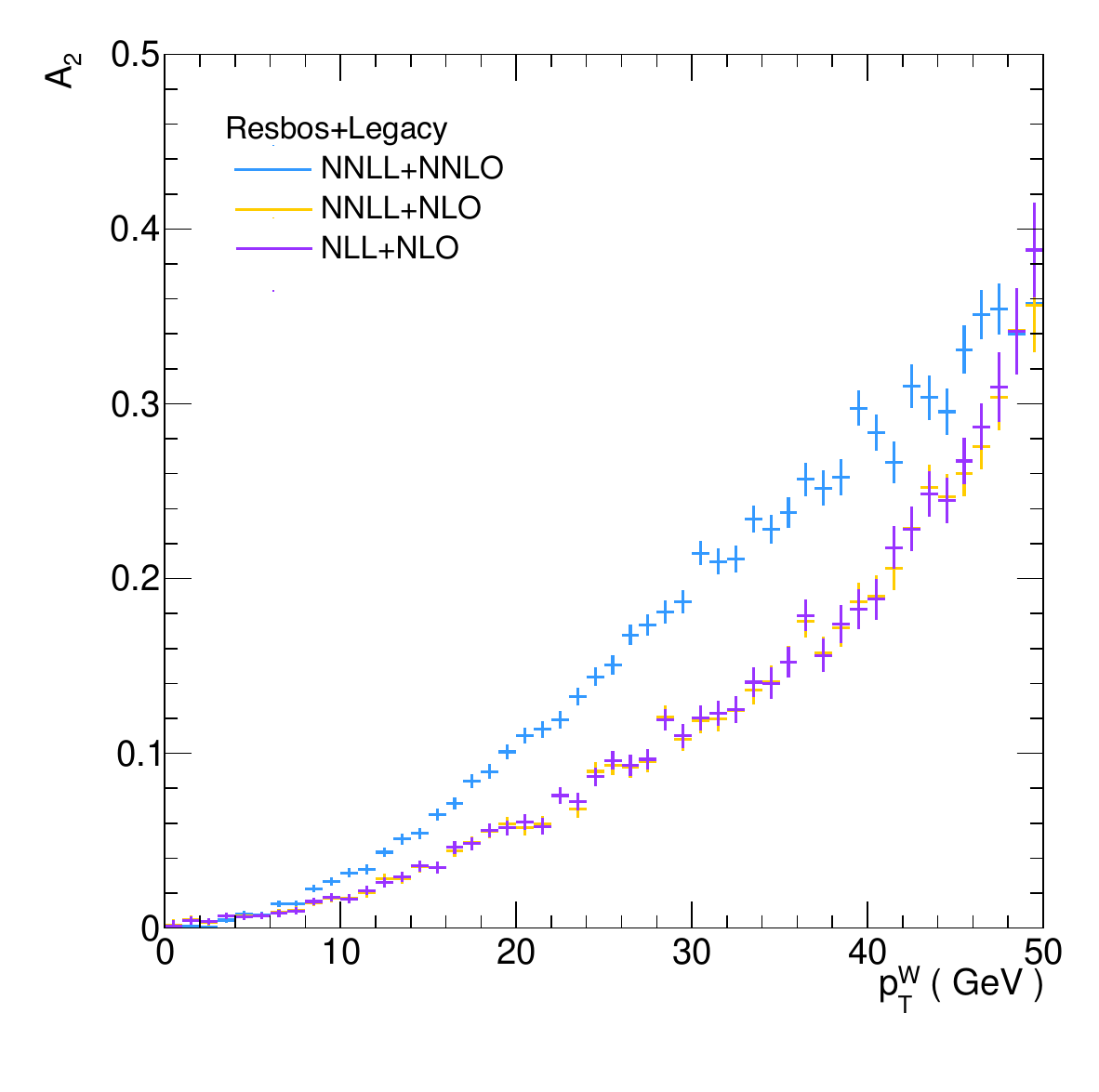}}
	\caption{$A_0$ and $A_2$ distributions with the dependence of $p_{T}^{W}$, generated by {\sc ResBos2} at different accuracy reyling on the $Y$ piece provided the {\sc Legacy} code. In fact, no differences between NLO and NNLO calculations are expected.}
	\label{fig:resbos2ai}
\end{figure}

Given that not all necessary components for the production of the original setups are available anylonger, the following  basic assumptions had to be made. First, since the original input file used by the old version of {\sc ResBos} to generate the default background prediction is unavailable, we regenerated it using {\sc ResBos2} at the same NLL+NLO accuracy. Notably, similar issues with the $A_i$ angular coefficients can be observed across different accuracy levels of {\sc ResBos2} predictions, as shown in Fig.~\ref{fig:resbos2ai}. This allows us to still estimate the impact of angular mismodeling. Second, since the {\sc CTEQ6M} PDF set is no longer available through {\sc LHAPDF}, we use {\sc CTEQ6.1} instead.

It is worthwhile to understand that {\sc ResBos} requires two components for generating particle-level events for the Drell–Yan process. The first is {\sc Legacy}, which provides two types of input grid files. The second is {\sc ResBos}, which reads these grid files and generates the events. The two grid file types correspond to the $W$ piece, which relates to the order of the resummation, and the $Y$ piece, which pertains to the order of the fixed-order calculation. At present, the newer {\sc ResBos2} package can generate only the $W$ piece. Therefore, predictions from {\sc ResBos2} are produced by combining the $W$ piece from {\sc ResBos2} with the $Y$ piece generated by {\sc Legacy}.

In Ref.\cite{LHC-TeVMWWorkingGroup:2023zkn}, a comparison was made between {\sc ResBos2} at NNLL+NNLO accuracy and {\sc ResBos} at NLL+NLO accuracy. This led to the conclusion that the issue with the angular coefficients was resolved in the {\sc ResBos2} prediction at NNLL+NNLO accuracy, which is indeed correct. However, as shown in Fig.\ref{fig:resbos2ai}, the root of this issue lies in the difference between the NLO and NNLO calculations in the $Y$ piece provided by the {\sc Legacy} code. Actually, the angular coefficients at NLO should be consistent with those at NNLO as shown in Fig.~6 of Ref.~\cite{LHC-TeVMWWorkingGroup:2023zkn} by {\sc DYNNLO}, which indicates that the angular coefficients are not correctly resummed by {\sc Legacy} at NLO. This implies that the angular coefficient problem can still be identified by comparing {\sc ResBos2} predictions at NNLL+NLO and NNLL+NNLO accuracies. We are therefore confident that our assumptions hold and provide an accurate description of the $Z$ boson production that was originally used.

%One is called {\sc Legacy}, which is used to provide the two kinds of input grid files, the other is called {\sc ResBos}, which reads the input grid files and generates the events. The two kinds of input grid files are called the $W$ piece, which is related to the order of resummation, and the $Y$ piece, which is related to the order of the fixed order calculation. Currently, the package called {\sc ResBos2} can only generate the $W$ piece. Currently, the prediction of {\sc ResBos2} is generated with the $W$ piece from the package {\sc ResBos2} and the $Y$ piece from the package {\sc Legacy}. In Ref.\cite{LHC-TeVMWWorkingGroup:2023zkn}, Only {\sc ResBos2} at the accuracy of NNLL+NNLO and {\sc ResBos} at the accuracy of NLL+NLO were compared, then concluded that the problem of the angular coefficients were solved in the {\sc ResBos2} prediction at the accuracy of NNLL+NNLO which is true. However, from Fig.\ref{fig:resbos2ai}, this problem is caused by the difference between NLO and NNLO calculation, which is related to the $Y$ piece calculation from {\sc Legacy}. This means we can still find the problem by comparing the {\sc ResBos2} predictions at the accuracy of NNLL+NLO and NNLL+NNLO.

% Hi Matthias I added this part as footnote since I don't know where to put it.

%An overview of the generated samples, including a brief description of each, is presented in Table \ref{xxx}.

%\begin{figure}
%    \subfigure[]{\includegraphics[width = 0.32\textwidth]{ Electron_Signal_Mt_vs_Resummation.pdf}}
%    \subfigure[]{\includegraphics[width = 0.32\textwidth]{ Electron_Signal_Lepton_Pt_vs_Resummation.pdf}}
%    \subfigure[]{\includegraphics[width = 0.32\textwidth]{ Electron_Signal_Neutrino_Pt_vs_Resummation.pdf}}
%    \\
%    \subfigure[]{\includegraphics[width = 0.32\textwidth]{ Muon_Signal_Mt_vs_Resummation.pdf}}
%    \subfigure[]{\includegraphics[width = 0.32\textwidth]{ Muon_Signal_Lepton_Pt_vs_Resummation.pdf}}
%    \subfigure[]{\includegraphics[width = 0.32\textwidth]{ Muon_Signal_Neutrino_Pt_vs_Resummation.pdf}}
%    \caption{The $M_T$(a/d), $p_T^\ell$(b/e) and $p_T^\nu$(c/f) distributions from the simulated signal samples with different resummation calculations. The top three distributions are for the electron channel and the bottom three distributions are for the muon channel.}
%\end{figure}

%%%%%%%%%%%%%%%%%%%%%%%%%%%%%%%%%%%%%%%%%%%%%%%%%%%%%%%%%%%%%%%%%%%%

\subsection{Cross Section Predictions and Acceptance Effects}
\label{ssec:xsec_and_acc}

In the CDF measurement, the total number of $Z\rightarrow \ell\ell$ background events is normalized using its fraction relative to the signal, denoted by $f$, as defined in Eq.\ref{eq:frac}.
\begin{equation}
	\label{eq:frac}
	f = \frac{A_Z \times \sigma_Z}{A_W \times \sigma_W} = \frac{1}{A_{W/Z} \times R_{W/Z}} 
\end{equation}
Here, $R_{W/Z} = \frac{\sigma_W}{\sigma_Z}$ represents the ratio of the total cross sections for $W$ and $Z$ boson production. The $Z$ boson cross section, $\sigma_Z$, is computed within the invariant mass window of 66 to 116GeV. The other term, $A_{W/Z}$, refers to the ratio of the acceptances for the two processes.

By default, $R_{W/Z}$ is calculated using {\sc FEWZ} \cite{Melnikov:2006kv, Gavin:2010az} at N$^3$LL+N$^3$LO accuracy with the {\sc MSTW2008} PDF set, yielding a value of 10.96, in agreement with the CDF estimation. We first validated this result using {\sc DYTurbo} \cite{DYTurbo} at the same theoretical accuracy and with the same PDF set. We then studied the impact of using different PDF sets on the value of $R_{W/Z}$. The total cross sections for both processes, along with their ratio $R_{W/Z}$, are summarized in Table~\ref{tab:rwz_pdf}. The value of $R_{W/Z}$ obtained using the {\sc MSTW2008} PDF set is the largest among all considered PDF sets. This includes newer sets such as {\sc MMHT2014} and {\sc MSHT20}, which supersede {\sc MSTW2008}, as well as various versions from the {\sc NNPDF} and {\sc CTEQ} collaborations. When the PDF set used in the $R_{W/Z}$ calculation is varied, the resulting impact on the final \mW{} measurement is shown in Table~\ref{tab:rwz_mw}.

The largest shift is observed when using {\sc MSHT20}, which leads to a change of over 4~MeV in the muon channel \mT{} and over 5~MeV in the \pTl{} and \pTnu{} observables. All observed shifts are negative, indicating that a lower value of $R_{W/Z}$ results in a higher estimated contribution from $Z \rightarrow \ell\ell$ background events, thereby reducing the measured value of $m_W$. This behavior is consistent with expectations.
  
\begin{table}
\centering
\begin{tabular}{p{5.5cm}|p{1.5cm}p{1.5cm}p{1.5cm}|p{1.5cm}p{1.5cm}p{1.5cm}}
	\hline
	PDF & $m_{T}^{e}$ & $p_{T}^{e}$ & $p_{T}^{\nu}$ & $m_{T}^{\mu}$ & $p_{T}^{\mu}$ & $p_{T}^{\nu}$ \\
	\hline
	MSTW2008 nnlo& 0 & 0 & 0 & 0 & 0 & 0 \\
	MSHT20 an3lo & -0.2 & -0.3 & -0.3 & -4.1 & -5.1 & -5.2 \\
	MSHT20qed an3lo & -0.2 & -0.3 & -0.3 & -4.2 & -5.2 & -5.3\\
	\hline
    NNPDF3.1 nnlo & -0.1 & -0.2 & -0.2 & -2.5 & -3.1 & -3.2 \\
    NNPDF4.0 nnlo & -0.1 & -0.2 & -0.2 & -2.8 & -3.5 & -3.6\\
    \hline
    CTEQ6.1 & -0.2 & -0.2 & -0.2 & -3.0 & -3.7 & -3.8\\
    CT18 nnlo & -0.1 & -0.1 & -0.1 & -1.7 & -2.2 & -2.2 \\
    \hline
\end{tabular}
\caption{The differences between the measured $m_W$ values and the input $m_W$ value, $\Delta m_W = m_W^{(\text{obs})}- m_W^{(\text{in})}$, with the different $R_{W/Z}$ used to normalize the $Z\rightarrow \ell \ell$ background.} 
\label{tab:rwz_mw}
\end{table}

The other term, $A_{W/Z}$, is calculated using the default Monte Carlo samples generated by {\sc ResBos}, employing the CTEQ6M PDF set at NLL+NLO accuracy. Its impact is estimated by varying the PDF sets and the perturbative accuracy, with results summarized in Table~\ref{tab:acc}. It can be seen that the variation in $A_{W/Z}$ for the muon channel is on the order of $\mathcal{O}(10^{-3})$, which is negligible compared to the impact from $R_{W/Z}$.               

\begin{table}
	\centering
	\begin{tabular}{p{4.2cm}|p{4.2cm}|rrrrrr}
		\hline
		PDF & Model & $A_{W}^{e}$ & $A_{Z}^{e}$ & $A_{W/Z}^{e}$ & $A_{W}^{\mu}$ & $A_{Z}^{\mu}$ & $A_{W/Z}^{\mu}$ \\
	    \hline
	    %NNPDF3.1 nnlo & N$^3$LL+NNLO & 18.48\% & 0.33\% & 56.00 & 18.78\% & 19.50\% & 0.963 \\
	    %NNPDF3.1 nnlo & W432+y  & 18.73\% & 0.34\% & 55.09 & 19.04\% & 19.81\% & 0.961 \\
	    NNPDF3.1 nnlo & NNLL+NNLO & 18.7\% & 0.3\% & 54.9 & 19.0\% & 19.7\% & 0.96 \\
	    NNPDF3.1 nnlo & NNLL+NLO  & 18.9\% & 0.3\% & 55.6 & 19.2\% & 20.0\% & 0.96 \\
	    %NNPDF3.1 nnlo & W221+yk & 18.68\% & 0.34\% & 54.94 & 18.98\% & 19.72\% & 0.963 \\
	    NNPDF3.1 nnlo & NLL+NLO  & 18.9\% & 0.3\% & 55.7 & 19.3\% & 20.0\% & 0.96 \\
	    %NNPDF3.1 nnlo & W211+yk & 18.82\% & 0.34\% & 55.35 & 19.13\% & 19.89\% & 0.962 \\
	    %NNPDF3.1 nnlo & NLL'+NLO  & 19.09\% & 0.35\% & 54.54 & 19.40\% & 20.21\% & 0.960 \\
	    \hline
	    %CTEQ6.1       & N$^3$LL+NNLO & 18.40\% & 0.33\% & 55.76 & 18.70\% & 19.36\% & 0.966 \\
	    %CTEQ6.1       & W432+y  & 18.64\% & 0.34\% & 54.82 & 18.95\% & 19.67\% & 0.964 \\
	    CTEQ6.1       & NNLL+NNLO & 18.6\% & 0.3\% & 54.6 & 18.9\% & 19.6\% & 0.97 \\
	    CTEQ6.1       & NNLL+NLO  & 18.8\% & 0.3\% & 55.3 & 19.1\% & 19.9\% & 0.96 \\
	    %CTEQ6.1       & W221+yk & 18.60\% & 0.34\% & 54.71 & 18.90\% & 19.59\% & 0.965 \\
	    CTEQ6.1       & NLL+NLO  & 18.9\% & 0.3\% & 55.4 & 19.2\% & 19.9\% & 0.96 \\
	    %CTEQ6.1       & W211+yk & 18.75\% & 0.34\% & 55.15 & 19.05\% & 19.76\% & 0.964 \\
	    %CTEQ6.1       & NLL'+NLO  & 18.99\% & 0.34\% & 55.85 & 19.30\% & 20.07\% & 0.962 \\
	    \hline
	    %MSTW2008 nnlo & N$^3$LL+NNLO & 18.64\% & 0.34\% & 54.82 & 18.94\% & 19.56\% & 0.968 \\
	    %MSTW2008 nnlo & W432+y  & 18.88\% & 0.34\% & 55.53 & 19.19\% & 19.86\% & 0.967 \\
	    MSTW2008 nnlo & NNLL+NNLO & 18.8\% & 0.3\% & 55.3 & 19.1\% & 19.7\% & 0.97 \\
	    MSTW2008 nnlo & NNLL+NLO  & 19.1\% & 0.3\% & 56.0 & 19.4\% & 20.1\% & 0.97 \\
	    %MSTW2008 nnlo & W221+yk & 18.83\% & 0.34\% & 55.38 & 19.14\% & 19.78\% & 0.968 \\
	    MSTW2008 nnlo & NNLL+NLO  & 19.1\% & 0.3\% & 56.2 & 19.4\% & 20.1\% & 0.97 \\
	    %MSTW2008 nnlo & W211+yk & 18.98\% & 0.34\% & 55.82 & 19.29\% & 19.94\% & 0.968 \\
	   % MSTW2008 nnlo & NLL'+NLO  & 19.23\% & 0.35\% & 54.94 & 19.55\% & 20.25\% & 0.966 \\
	    \hline
	    %CT18 nnlo     & N$^3$LL+NNLO & 18.61\% & 0.33\% & 56.39 & 18.90\% & 19.48\% & 0.970 \\
	    %CT18 nnlo     & W432+y  & 18.86\% & 0.34\% & 55.47 & 19.17\% & 19.79\% & 0.969 \\
	    CT18 nnlo     & NNLL+NNLO & 18.8\% & 0.3\% & 55.2 & 19.1\% & 19.7\% & 0.97 \\
	    CT18 nnlo     & NNLL+NLO  & 19.0\% & 0.3\% & 56.0 & 19.4\% & 20.0\% & 0.97 \\
	    %CT18 nnlo     & W221+yk & 18.81\% & 0.34\% & 55.32 & 19.11\% & 19.70\% & 0.970 \\
	    CT18 nnlo     & NLL+NLO  & 19.1\% & 0.3\% & 56.1 & 19.4\% & 20.0\% & 0.97 \\
	    %CT18 nnlo     & W211+yk & 18.95\% & 0.34\% & 55.74 & 19.26\% & 19.87\% & 0.970 \\
	    %CT18 nnlo     & NLL'+NLO  & 19.22\% & 0.35\% & 54.91 & 19.54\% & 20.19\% & 0.968 \\
	    \hline
	\end{tabular}
    \caption{Acceptance of $p\bar{p}\rightarrow W^+ \rightarrow \ell^+ \nu$ and  $p\bar{p}\rightarrow Z/\gamma^*\rightarrow \ell^+ \ell^-$ processes, predicted with different calculation orders from {\sc ResBos2}  }\label{tab:acc}
\end{table}

\subsection{The Shape of the $Z\rightarrow \ell \ell$ Background}

The shape of the $Z\rightarrow \ell\ell$ background is estimated directly from the default Monte Carlo (MC) sample, which was generated using {\sc ResBos} at NLL+NLO accuracy with the {\sc CTEQ6M} PDF set.

%To provide an updated prediction for the background shape, a straightforward approach is to use a sample generated with {\sc ResBos2} at NNLL+NNLO accuracy, employing the {\sc NNPDF3.1} PDF set. The impact of this update is presented in Table~\ref{tab:shape}. The effect of changing the resummation order from NLL to NNLL results in a shift of -0.16~MeV on the muon channel \mW{} measured value from the \mT{} distribution,  -0.36~MeV from the \pTl{} distribution and -0.53~MeV from the \pTnu{} distribution; the effect of updating the fixed-order calculation from NLO to NNLO leads to a shift of -1.07~MeV on the muon channel \mW{} measured value from the \mT{} distribution,  +0.10~MeV from the \pTl{} distribution and -2.40~MeV from the \pTnu{} distribution; and the change of PDF set from {\sc CTEQ6.1} to {\sc NNPDF3.1} yields a shift of of +0.59~MeV on the muon channel \mW{} measured value from the \mT{} distribution,  +1.35~MeV from the \pTl{} distribution and +1.04~MeV from the \pTnu{} distribution. The total combined impact from these three factors is a shift of +0.18~MeV on the muon channel \mW{} measured value from the \mT{} distribution,  +1.38~MeV from the \pTl{} distribution and -1.10~MeV from the \pTnu{} distribution.
To update the background shape prediction, we use a sample generated with {\sc ResBos2} at NNLL+NNLO accuracy using the {\sc NNPDF3.1} PDF set. The resulting shifts in the muon channel \mW{} measurement are shown in Table~\ref{tab:shape}. Changing the resummation from NLL to NNLL shifts the result by -0.2, -0.4, and -0.5~MeV for the \mT{}, \pTl{}, and \pTnu{} distributions, respectively. Updating the fixed-order calculation from NLO to NNLO gives shifts of -1.1, +0.1, and -2.4~MeV. Replacing the {\sc CTEQ6.1} PDF with {\sc NNPDF3.1} shifts the values by +0.6, +1.4, and +1.0~MeV. The total combined effect is +0.2~MeV for \mT{}, +1.4~MeV for \pTl{}, and -1.1~MeV for \pTnu{}.

\begin{figure}
	\subfigure[]{\includegraphics[width = 0.32\textwidth]{ 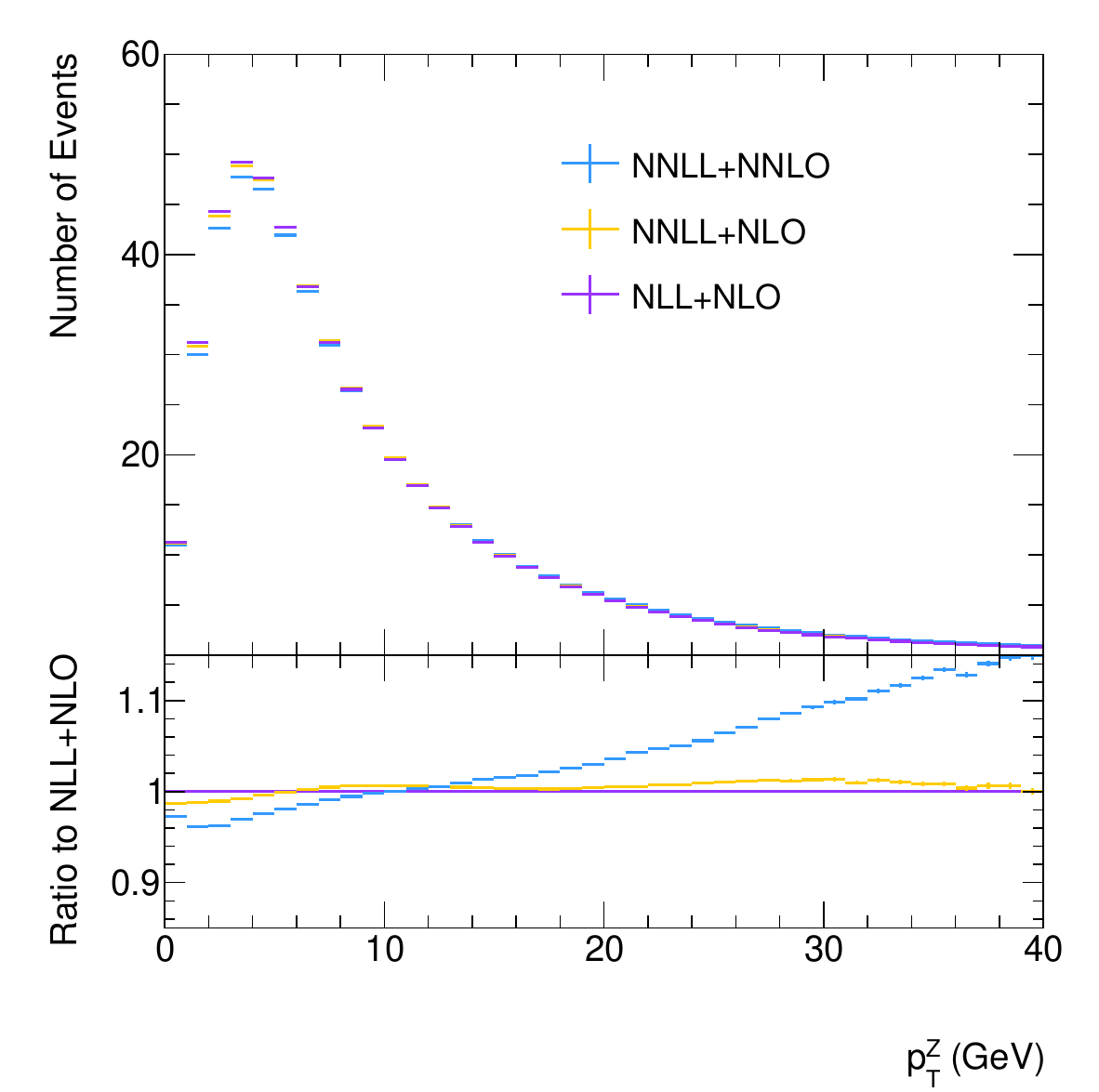}} 
	\subfigure[]{\includegraphics[width = 0.32\textwidth]{ 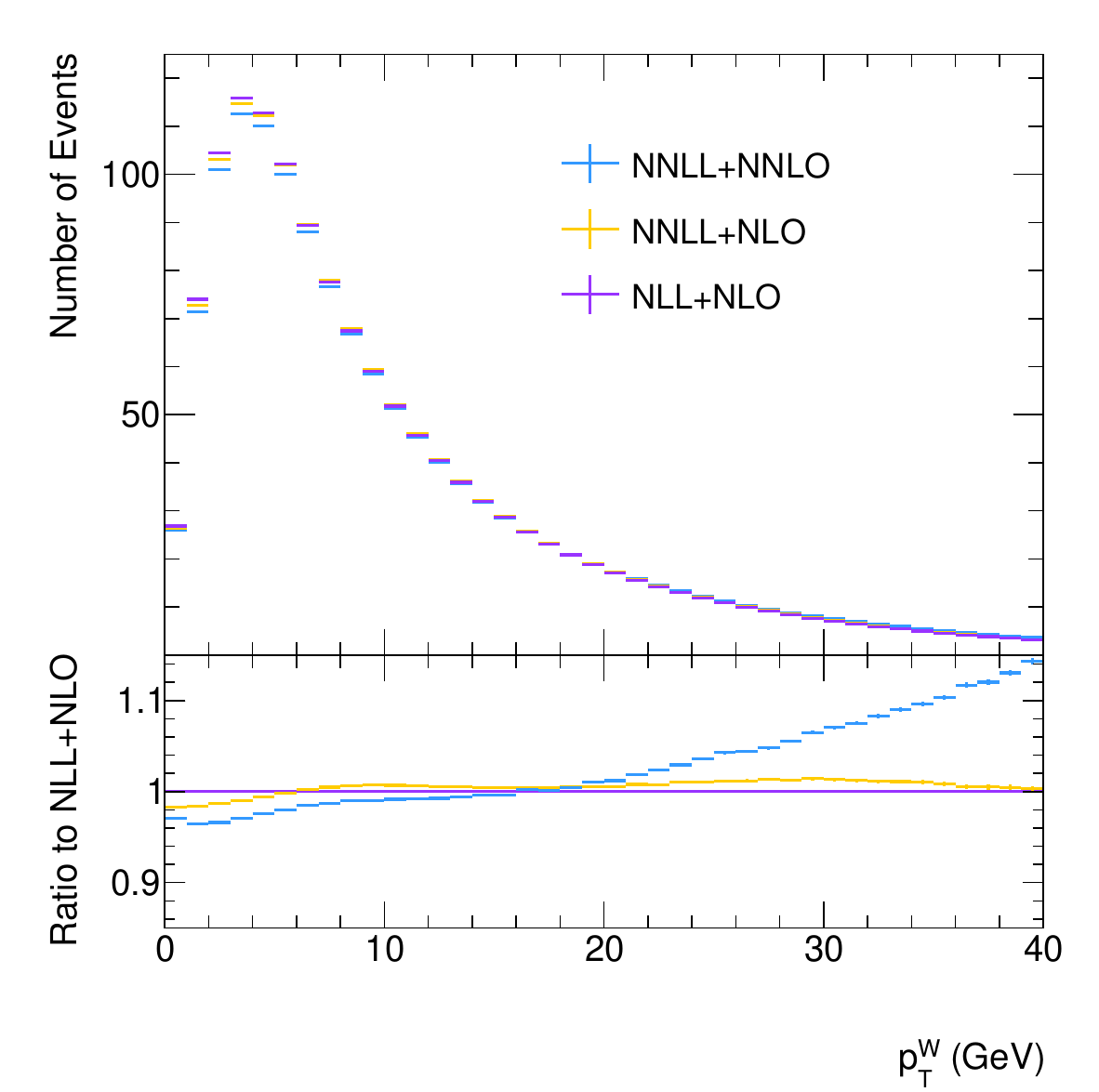}}
	\subfigure[]{\includegraphics[width = 0.32\textwidth]{ 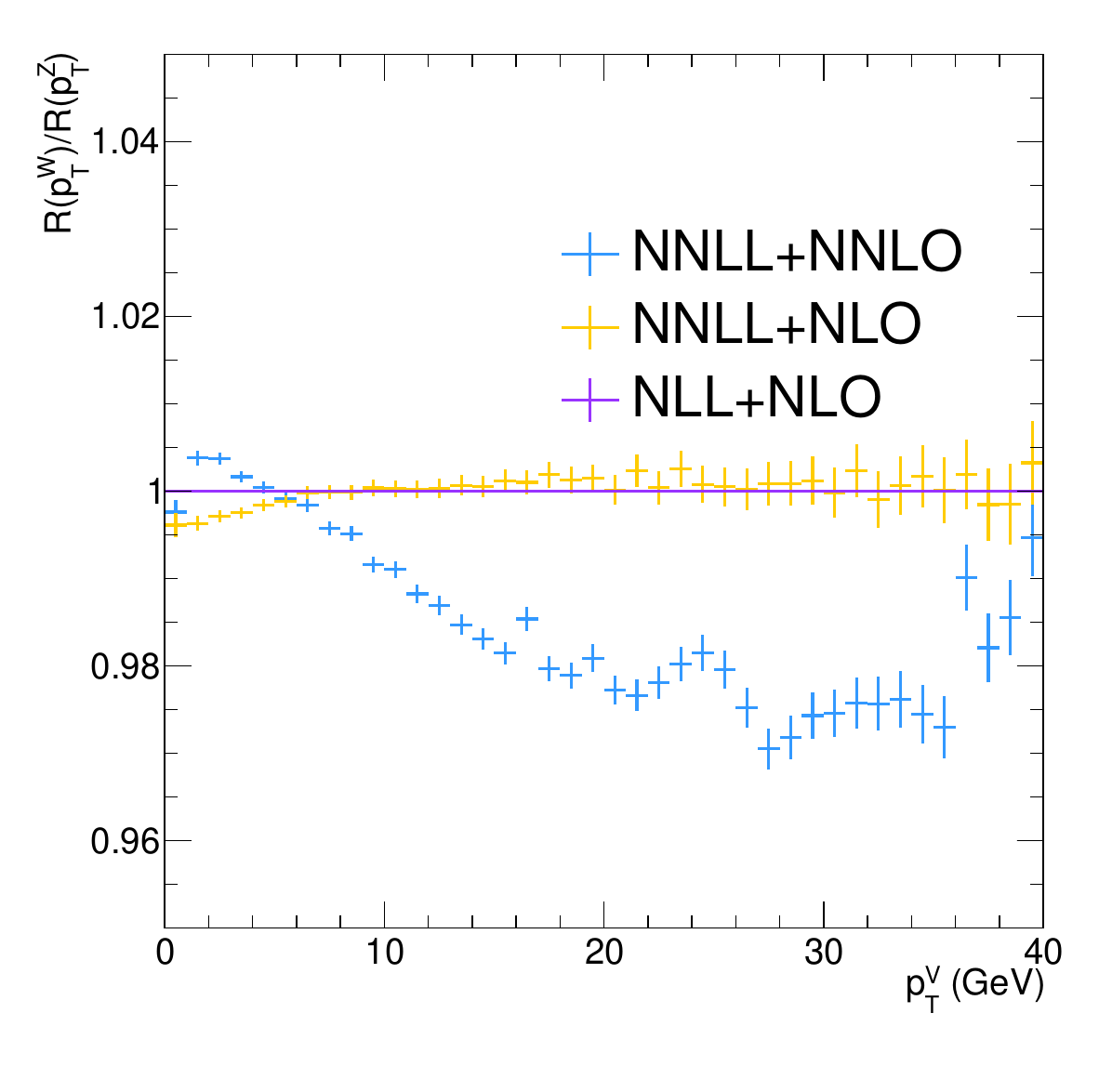}}
	\caption{$p_T^Z$ and $p_T^W$ distributions provided by {\sc ResBos2} with NLL+NLO, NNLL+NLO and NNLL+NNLO calculations are shown in (a) and (b), the non-perturbative parameters are not changed. The bottom panels of (a) and (b) show the ratios, $R(p_T^V)$, between the two distributions in the corresponding top panel with the dependence of $p_T^Z$(a) or $p_T^W$(b). The double ratio, $R(p_T^W)/R(p_T^Z)$, is shown in (c).} 
	\label{fig:ptv_resum}
\end{figure}

\begin{figure}
	\subfigure[]{\includegraphics[width = 0.32\textwidth]{ 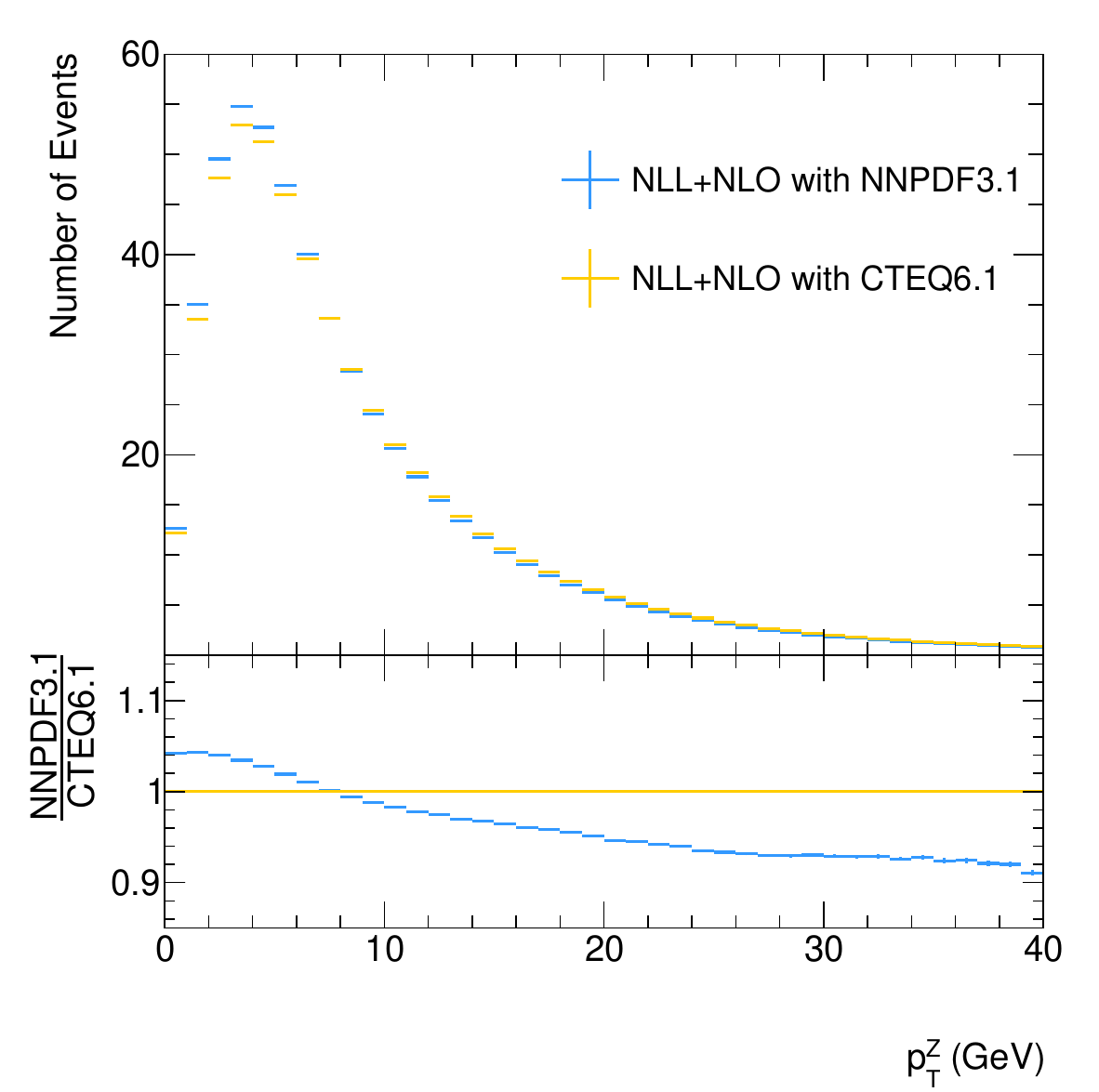}} 
	\subfigure[]{\includegraphics[width = 0.32\textwidth]{ 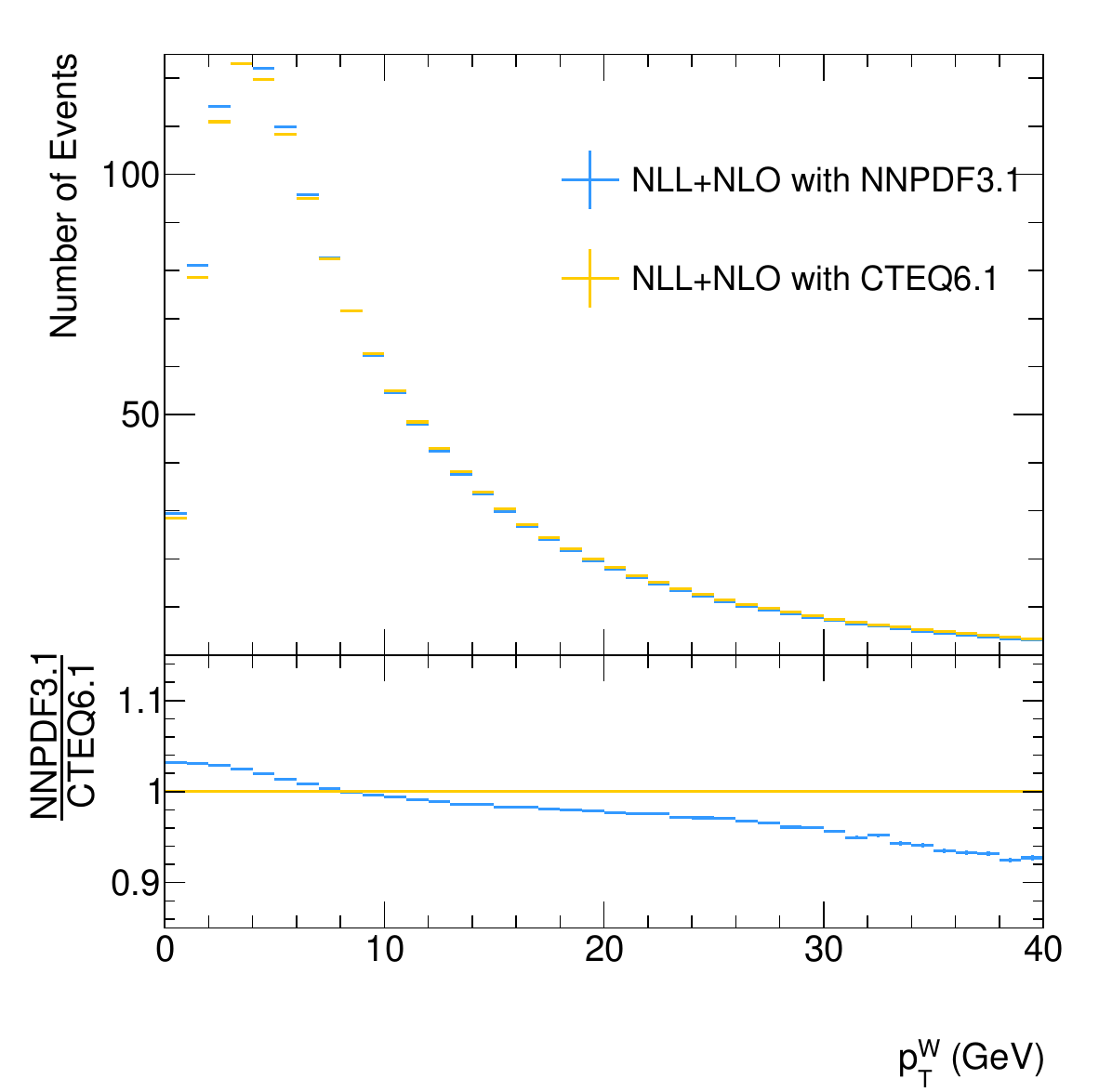}}
	\subfigure[]{\includegraphics[width = 0.32\textwidth]{ 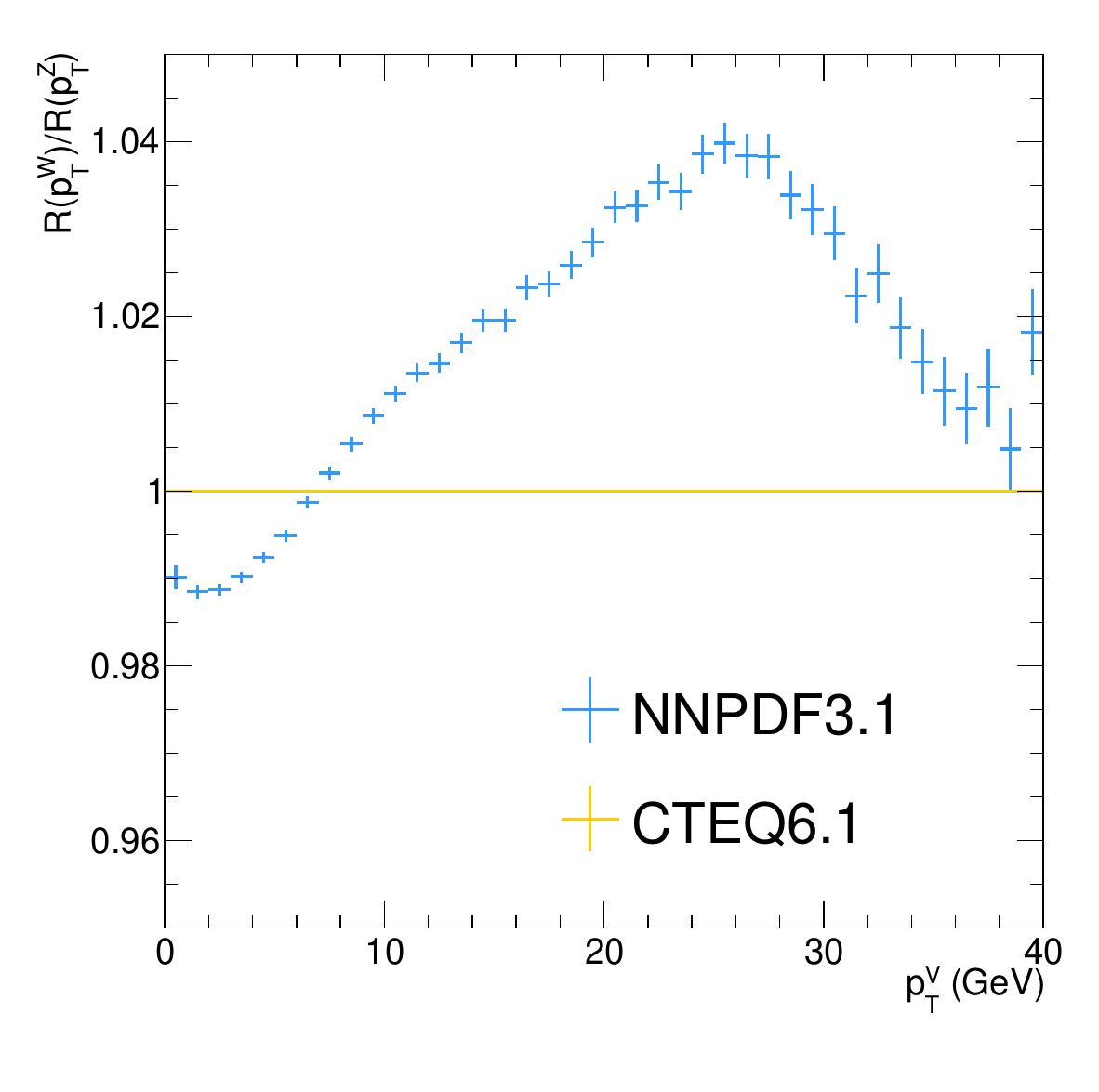}}
	\caption{$p_T^Z$ and $p_T^W$ distributions with CTEQ6.1 and NNPDF3.1 are shown in (a) and (b), the non-perturbative parameters are not changed. The bottom panels of (a) and (b) show the ratios, $R(p_T^V)$, between the two distributions in the corresponding top panel with the dependence of $p_T^Z$(a) or $p_T^W$(b). The double ratio, $R(p_T^W)/R(p_T^Z)$, is shown in (c).} 
	\label{fig:ptv_pdf}
\end{figure}

The other critical aspect is the modeling of the $p_T^V$ distribution, which can significantly affect the measured value of \mW{}, particularly in the low $p_T^V$ region. Both the perturbative QCD order and the choice of PDF set influence the shape of this distribution. The effect of the QCD order on the $p_T^V$ spectrum is shown in Fig.\ref{fig:ptv_resum}, while the corresponding impact of different PDF sets is illustrated in Fig.\ref{fig:ptv_pdf}.


In the low $p_T^V$ region, the shape of the distribution is dominated by non-perturbative QCD effects. These effects require a non-perturbative function with several free parameters. In {\sc ResBos}, this function is called BLNY\cite{RunITune}, which has three free parameters: $g_1$, $g_2$, and $g_3$. These parameters are tuned to match the shape of the $p_T^Z$ distribution. In the CDF measurement, the non-perturbative parameter values were tuned using the CTEQ6M PDF set with NLL+NLO accuracy. Therefore, in principle, these parameters should be retuned to account for the differences shown in Fig.\ref{fig:ptv_resum} and Fig.\ref{fig:ptv_pdf}.

However, in practice, this retuning is often replaced by a reweighting procedure. For example, in Ref.~\cite{LHC-TeVMWWorkingGroup:2023zkn}, when the PDF set of the signal sample was updated from {\sc CTEQ6M} to {\sc NNPDF3.1}, a reweighting of $p_T^W$ in the fiducial region was performed to preserve the $p_T^W$ modeling. Similarly, when updating the background $Z\rightarrow \ell \ell$ sample, a $p_T^Z$ reweighting can be applied to maintain the distribution unchanged.

After applying this $p_T^Z$ reweighting, the results from Table~\ref{tab:shape} are updated, as shown in Table~\ref{tab:shape_ptz}. From this updated table, the impact due to changing the resummation calculation from NLL to NNLL is a shift of -0.2~MeV on the muon channel \mW{} measured value from the \mT{} distribution,  -0.4~MeV from the \pTl{} distribution and -0.4~MeV from the \pTnu{} distribution, the impact due to changing the fixed-order calculation from NLO to NNLO is a shift of -3.6~MeV on the muon channel \mW{} measured value from the \mT{} distribution,  -3.1~MeV from the \pTl{} distribution and -5.3~MeV from the \pTnu{} distribution, and the impact due to changing the PDF set from {\sc CTEQ6.1} to {\sc NNPDF3.1} is a shift of +1.4~MeV on the muon channel \mW{} measured value from the \mT{} distribution,  +2.1~MeV from the \pTl{} distribution and +1.1~MeV from the \pTnu{} distribution. The total combined impact from all these factors is a shift of -1.2~MeV on the muon channel \mW{} measured value from the \mT{} distribution,  -0.4~MeV from the \pTl{} distribution and -3.2~MeV from the \pTnu{} distribution.

\begin{table}
		\centering
		\begin{tabular}{p{4.8cm}|p{4.8cm}|rrr|rrr}
		\hline
		PDF & Model & $m_{T}^{e}$ & $p_{T}^{e}$ & $p_{T}^{\nu}$ & $m_{T}^{\mu}$ & $p_{T}^{\mu}$ & $p_{T}^{\nu}$ \\
		\hline
		CTEQ6.1 & NLL+NLO & 0 & 0 & 0 & 0 & 0 & 0 \\
		CTEQ6.1 & NNLL+NLO & $<$0.1 & $<$0.1 & $<$0.1 & -0.2 & -0.4 & -0.5 \\
		CTEQ6.1 & NNLL+NNLO & $<$0.1 & $<$0.1 & $<$0.1 & -1.2 & -0.3 & -2.9 \\
		\hline
		NNPDF3.1 nnlo & NLL+NLO & $<$0.1 &$<$0.1 & $<$0.1 & 0.6 & 1.4 & 1.0 \\
		NNPDF3.1 nnlo & NNLL+NLO & $<$0.1 & $<$0.1 & 0.1 & 0.4 & 0.7 & 0.8 \\
		NNPDF3.1 nnlo & NNLL+NNLO & $<$0.1 & $<$0.1 & 0.1 & 0.2 & 1.4 & -1.1 \\
		\hline
			\end{tabular}
            \caption{The differences between the measured $m_W$ values and the input $m_W$ value, $\Delta m_W = m_W^{(\text{obs})}- m_W^{(\text{in})}$, with the different shapes of the $Z\rightarrow \ell \ell$ background predicted from different PDFs and different calculations.}
            \label{tab:shape} 
	\end{table}
	
	\begin{table}
		\centering
		\begin{tabular}{p{4.8cm}|p{4.8cm}|rrr|rrr}
			\hline
			PDF & Model & $m_{T}^{e}$ & $p_{T}^{e}$ & $p_{T}^{\nu}$ & $m_{T}^{\mu}$ & $p_{T}^{\mu}$ & $p_{T}^{\nu}$ \\
			\hline
			CTEQ6.1 & NLL+NLO & 0 & 0 & 0 & 0 & 0 & 0 \\
			CTEQ6.1 & NNLL+NLO & $<$0.1 & $<$0.1 & $<$0.1 & -0.2 & -0.4 & -0.4 \\
			CTEQ6.1 & NNLL+NNLO & -0.2 & -0.2 & -0.2 & -3.6 & -3.1 & -5.3 \\
			\hline
			NNPDF3.1 nnlo & NLL+NLO & $<$0.1 & 0.2 & $<$0.1 & 1.4 & 2.1 & 1.1 \\
			NNPDF3.1 nnlo & NNLL+NLO & $<$0.1 & 0.2 & 0.2 & 1.1 & 1.5 & 1.0 \\
			NNPDF3.1 nnlo & NNLL+NNLO & $<$0.1 & $<$0.1 & $<$0.1 & -1.2 & -0.4 & -3.2 \\
			%NNPDF3.1 nnlo & N$^3$LL+NNLO & 0.88 & 0.52 & 0.36 & 0.95 & 0.53 & 0.47 \\
			%NNPDF3.1 nnlo & N$^3$LL+NLO & 0.89 & 0.46 & 0.41 & 2.04 & -0.10 & 3.39 \\
			%NNPDF3.1 nnlo & NNLL+NNLO & 0 & 0 & 0 & 0 & 0 & 0 \\
			%NNPDF3.1 nnlo & NNLL+NLO & 0.13 & 0.17 & 0.16 & 2.31 & 1.82 & 4.15 \\
			%NNPDF3.1 nnlo & W221+yk & 0.91 & 0.44 & 0.39 & 1.61 & 1.28 & 2.65 \\
			%NNPDF3.1 nnlo & NLL+NLO & 0.14 & 0.20 & 0.09 & 2.59 & 2.43 & 4.21 \\
			%NNPDF3.1 nnlo & W211+yk & 0.78 & 0.36 & 0.26 & 2.00 & 2.06 & 2.74 \\
			%NNPDF3.1 nnlo & NLL'+NLO & 0.95 & 0.40 & 0.33 & 3.05 & 1.60 & 5.22 \\
			%\hline
			%CTEQ6.1 & N$^3$LL+NNLO & 0.85 & 0.50 & 0.29 & 0.10 & -1.14 & -1.20 \\
			%CTEQ6.1 & W432+y & 0.86 & 0.46 & 0.36 & 1.00 & -1.55 & 0.94 \\
			%CTEQ6.1 & NNLL+NNLO & -0.11 & -0.19 & -0.23 & -2.41 & -2.69 & -2.11 \\
			%CTEQ6.1 & NNLL+NLO & $<$0.01 & $<$0.01 & 0.03 & 1.00 & 0.02 & 2.73 \\
			%CTEQ6.1 & W221+yk & 0.86 & 0.47 & 0.25 & 0.70 & -0.29 & 0.34 \\
			%CTEQ6.1 & NLL+NLO & 0.05 & 0.02 & $<$0.01 & 1.22 & 0.38 & 3.15 \\
			\hline
	\end{tabular}
\caption{The differences between the measured $m_W$ values and the input $m_W$ value, $\Delta m_W = m_W^{(\text{obs})}- m_W^{(\text{in})}$, with the different shapes of the $Z\rightarrow \ell \ell$ background predicted from different PDFs and different calculations. Additional $p_T^Z$ reweighting is applied} 
\label{tab:shape_ptz}
\end{table}
\begin{figure}
	\subfigure[]{\includegraphics[width = 0.32\textwidth]{ 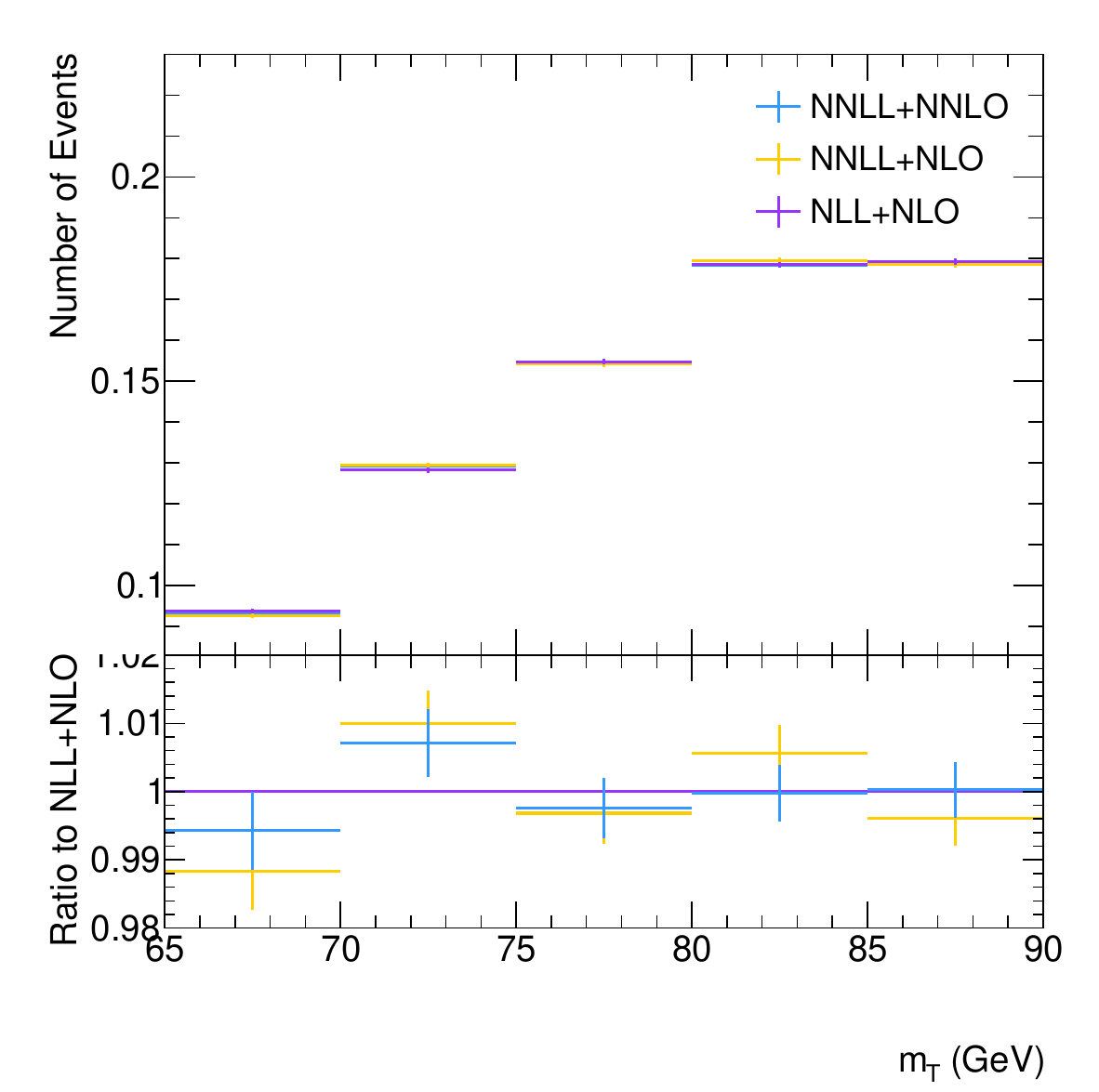}}
	\subfigure[]{\includegraphics[width = 0.32\textwidth]{ 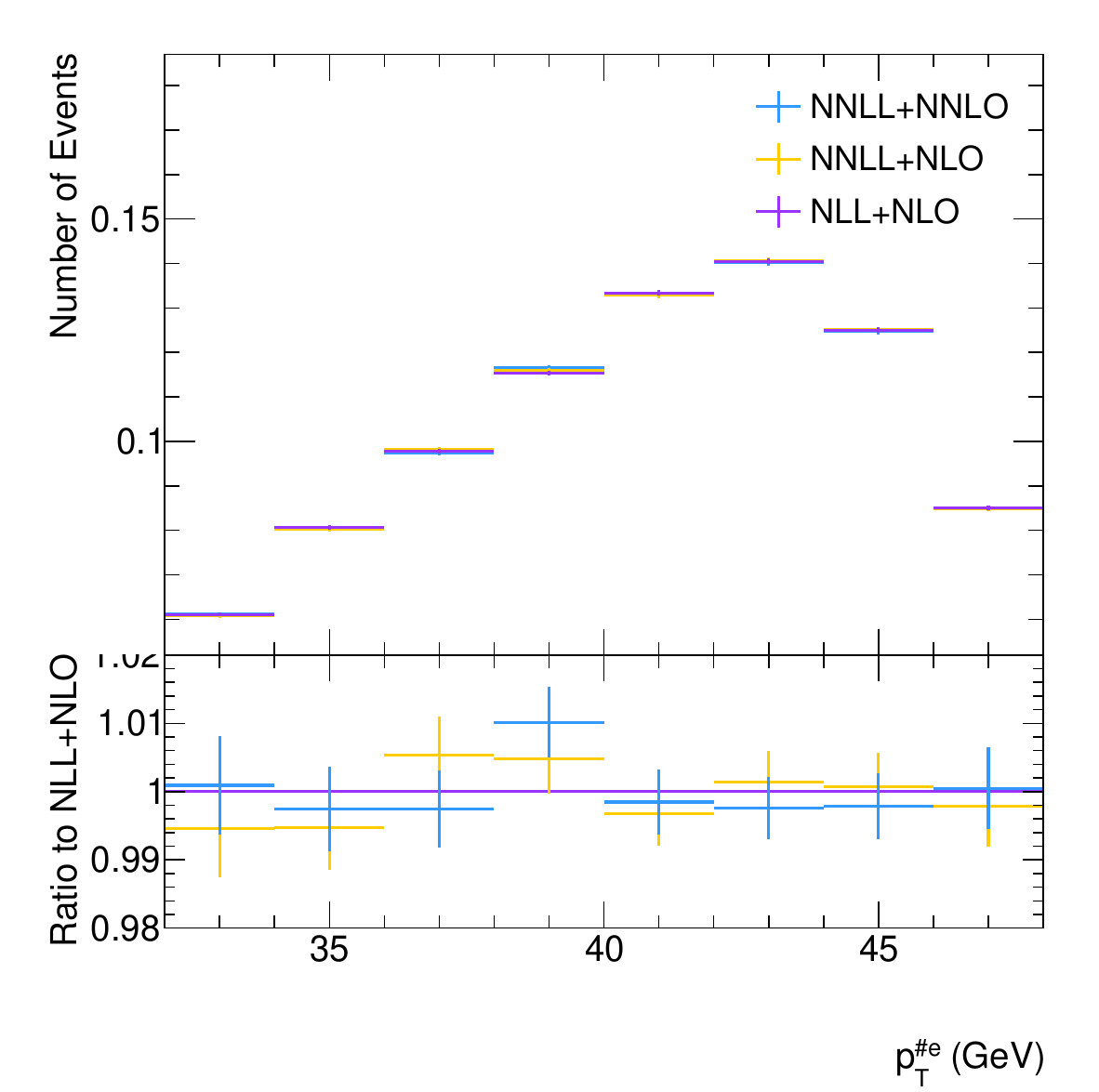}}
	\subfigure[]{\includegraphics[width = 0.32\textwidth]{ 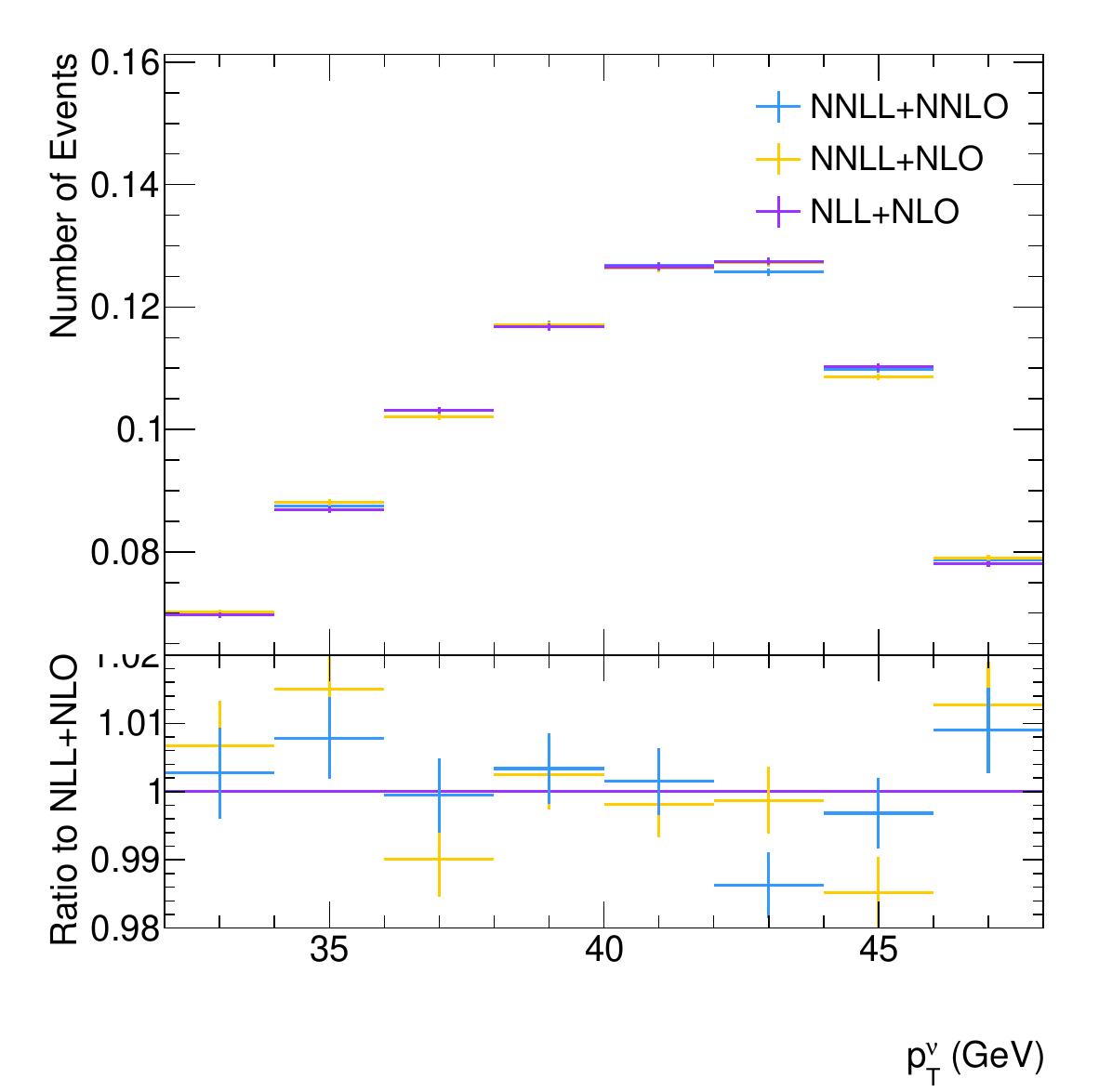}}
	\\
	\subfigure[]{\includegraphics[width = 0.32\textwidth]{ 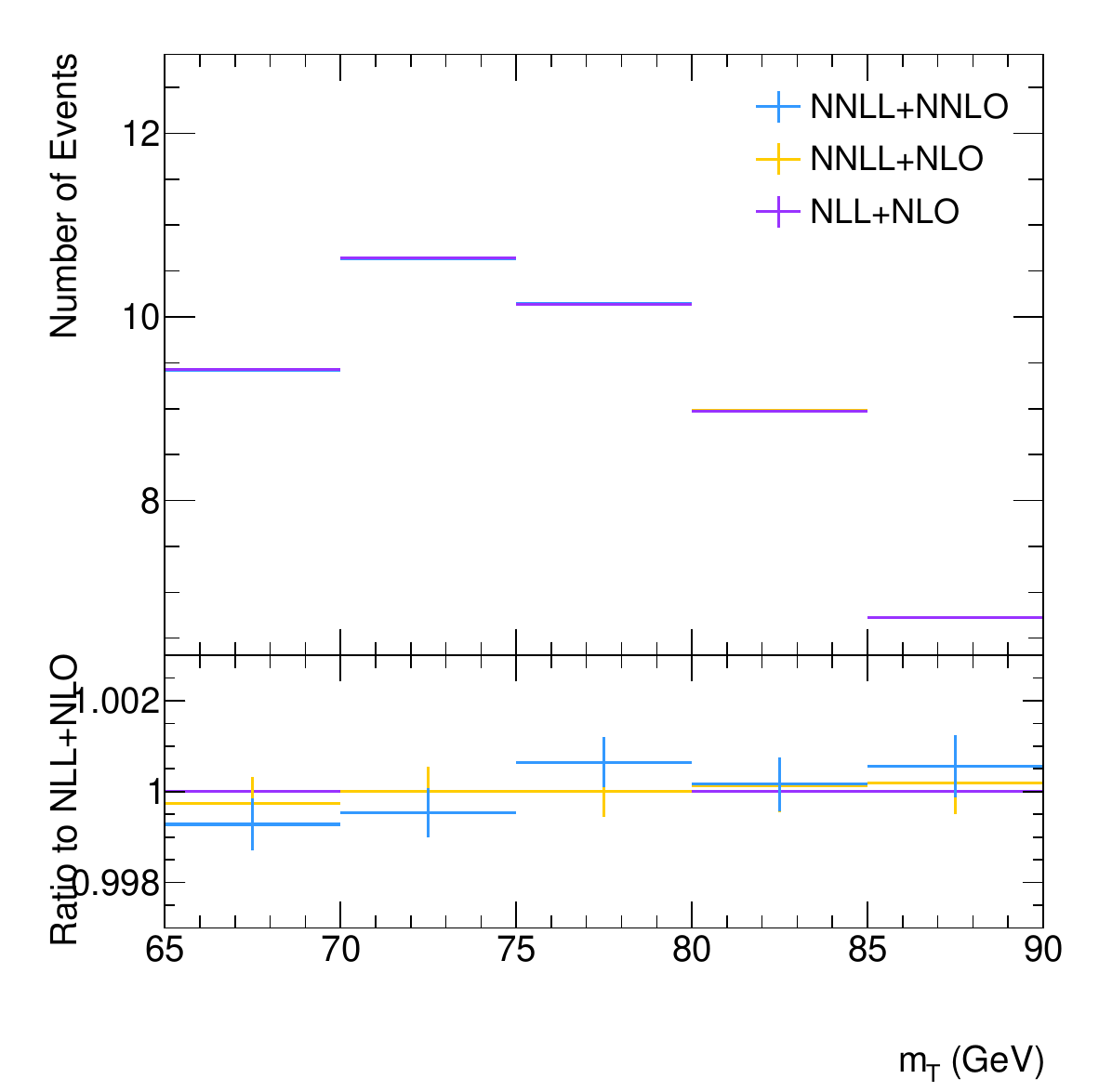}}
	\subfigure[]{\includegraphics[width = 0.32\textwidth]{ 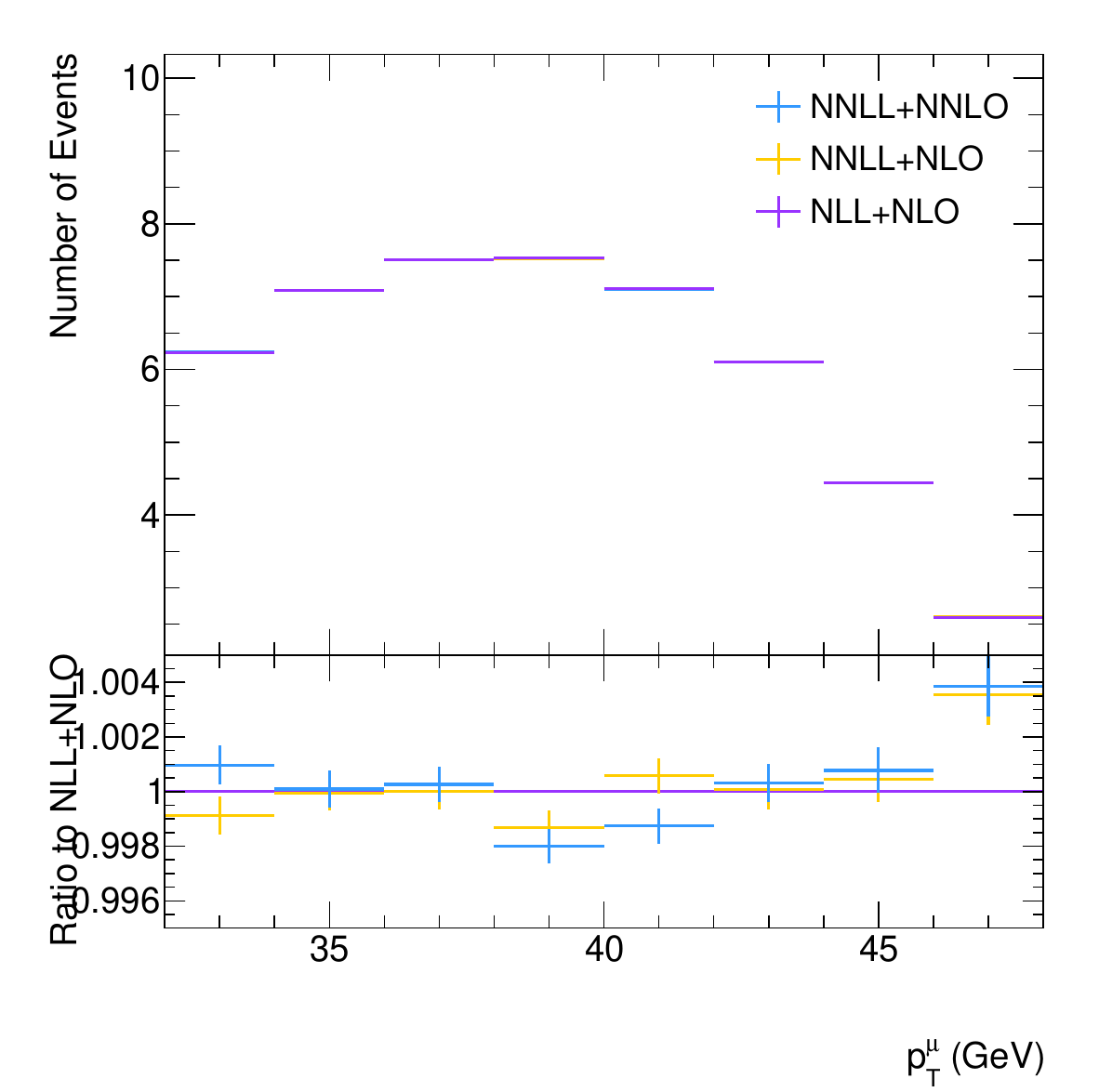}}
	\subfigure[]{\includegraphics[width = 0.32\textwidth]{ 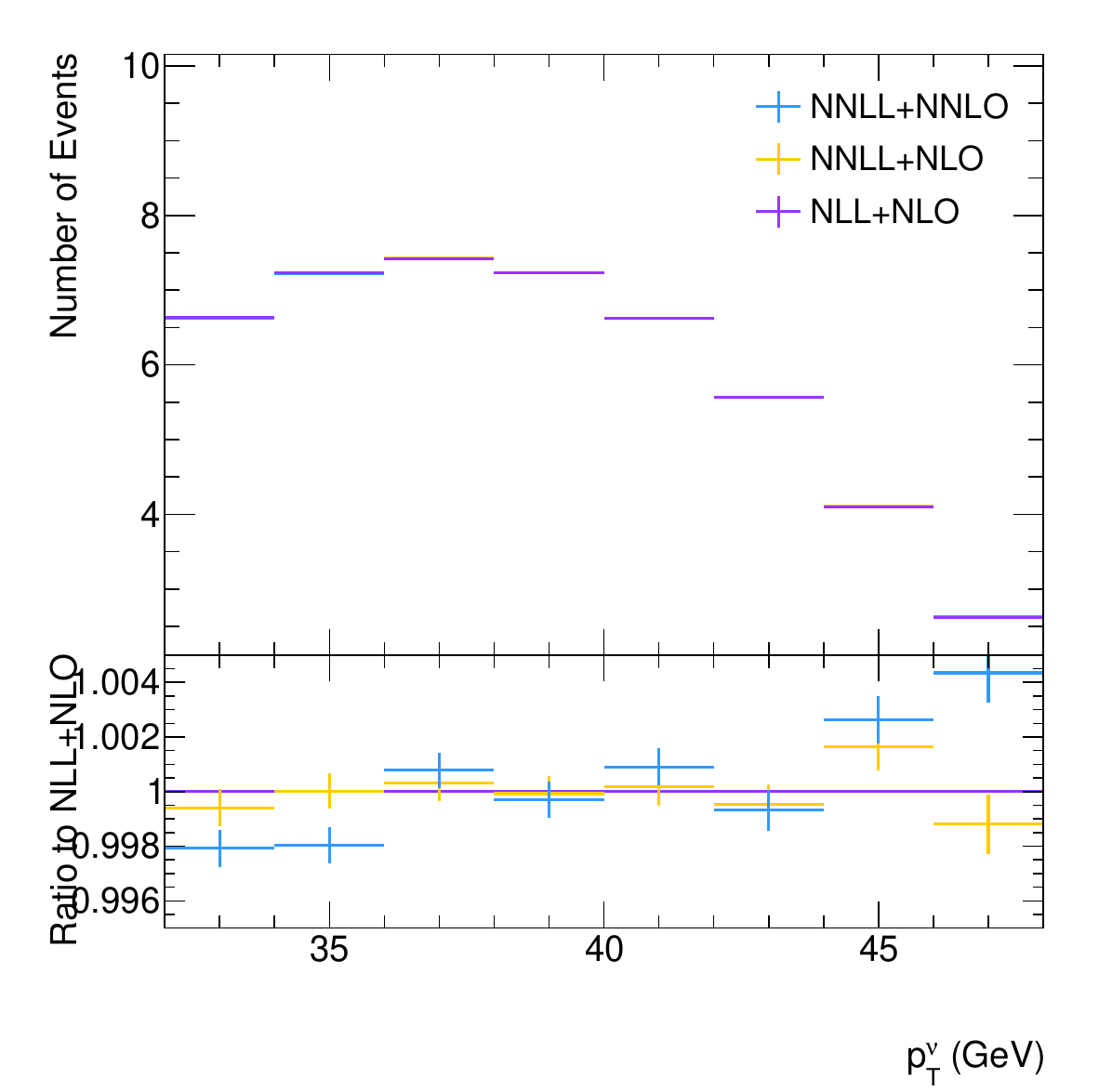}}
	\caption{The $M_T$(a/d), $p_T^\ell$(b/e) and $p_T^\nu$(c/f) distributions from the simulated background samples with different resummation calculations. The top three distributions are for the electron channel and the bottom three distributions are for the muon channel.}
\end{figure}
\begin{figure}
	\centering
	\subfigure[]{\includegraphics[width = 0.32\textwidth]{ 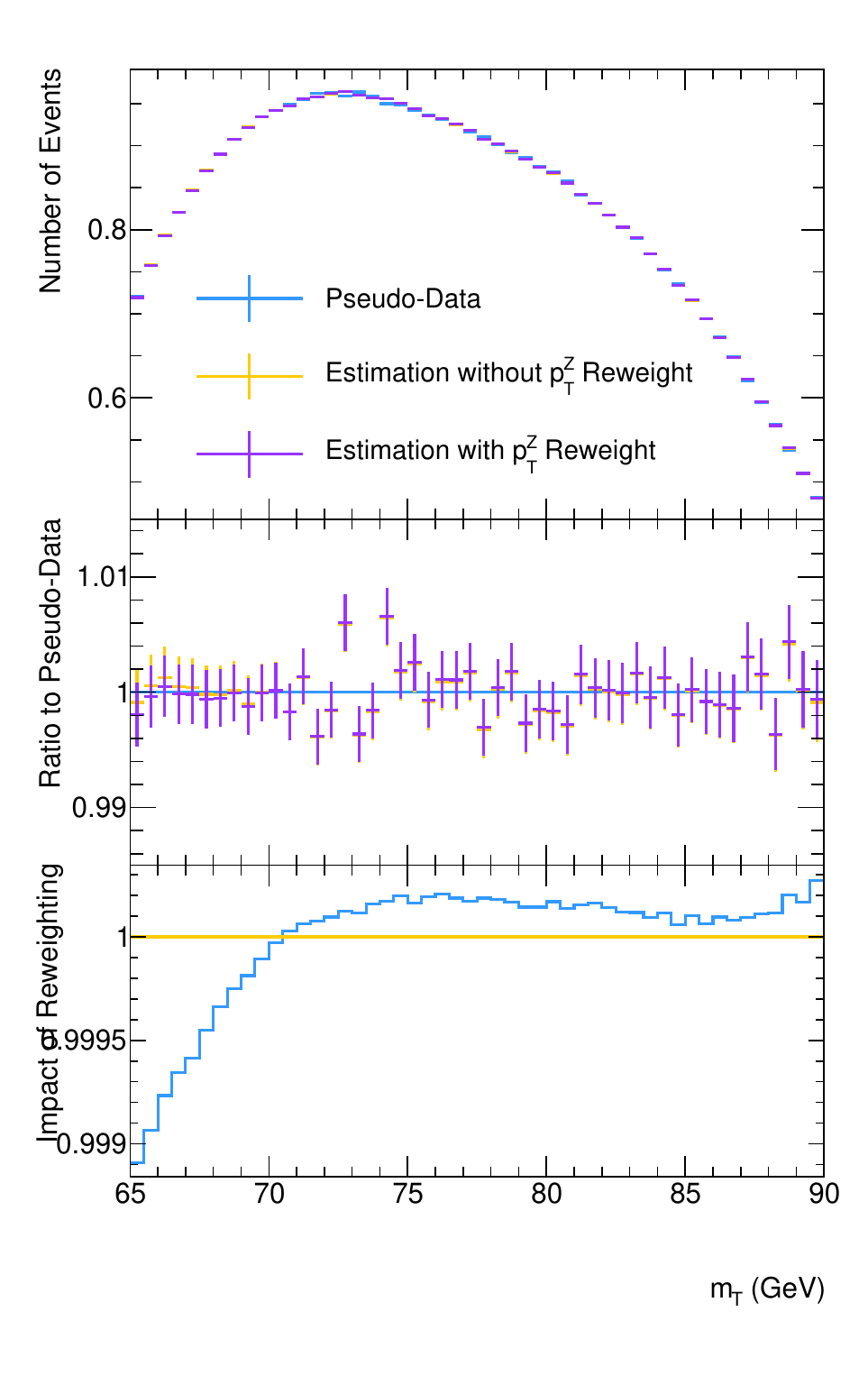}}
    \subfigure[]{\includegraphics[width = 0.32\textwidth]{ 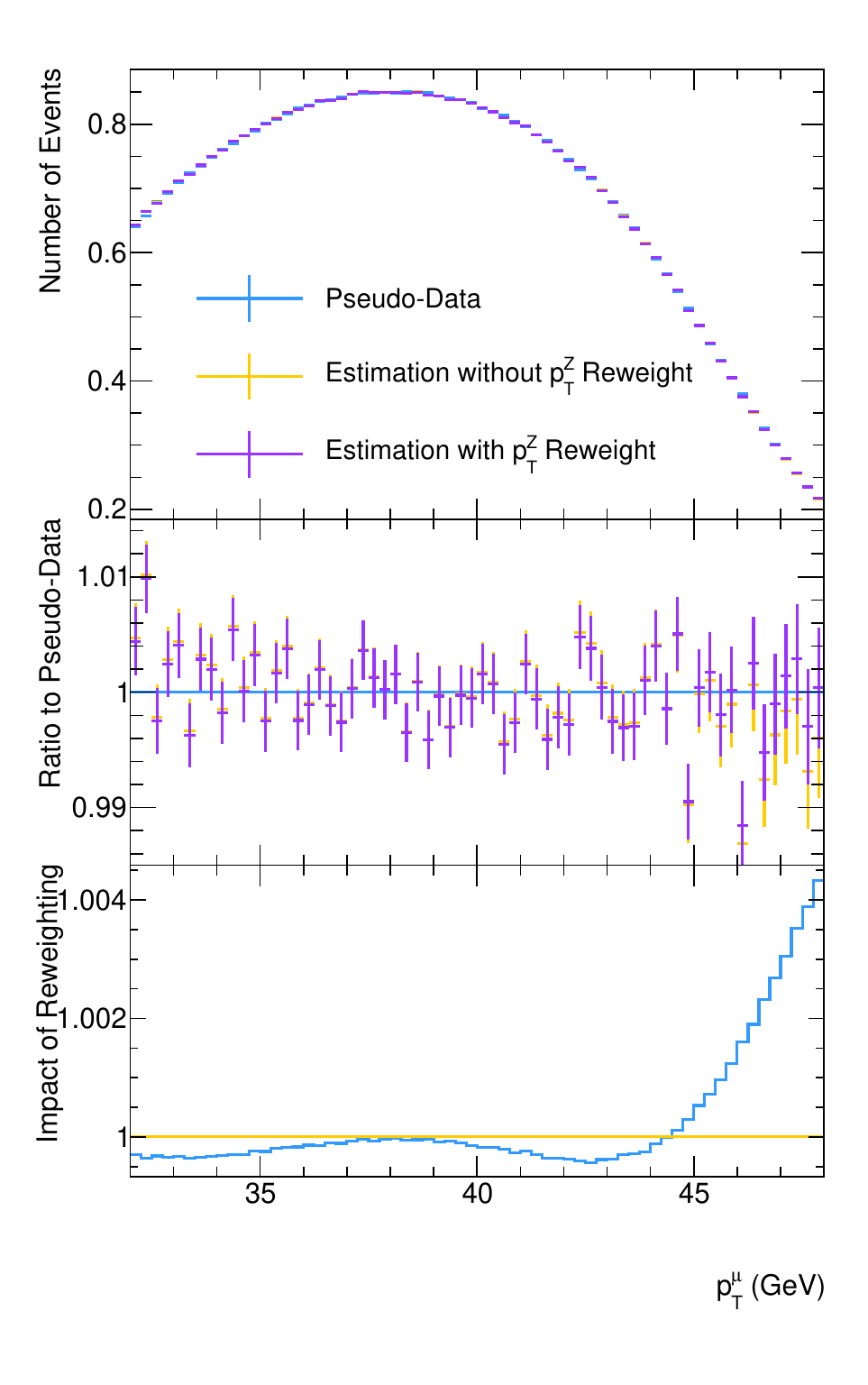}}
    \subfigure[]{\includegraphics[width = 0.32\textwidth]{ 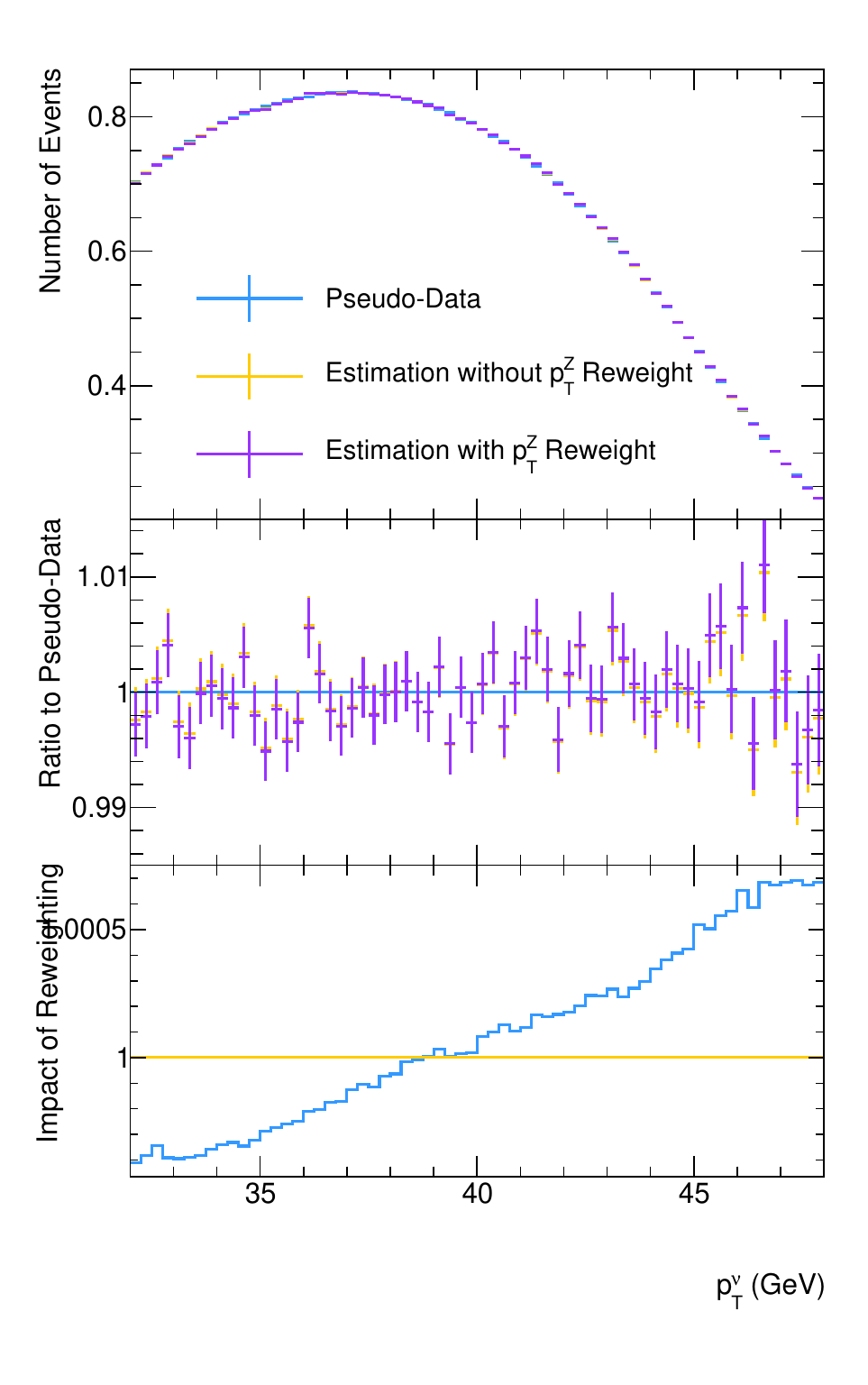}}
    \caption{Background distributions in pseudo-data generated by {\sc ResBos2} at the accuracy of NLL+NLO  with {\sc CTEQ6.1} and the estimation generated by {\sc ResBos2} at the accuracy of NNLL+NNLO with {\sc NNPDF 3.1} with or without its $p_T^Z$ distribution reweighted to that of the pseudo-data are shown in the top panel. In the middle, the comparision to the pseudo-data is shown and , in the bottom panel, the impact due to the $p_T^Z$ reweighting is shown. }
\end{figure}

\subsection{Overall Uncertainties}

Finally, a comprehensive test was conducted to evaluate the overall impact of updating the $Z\rightarrow \ell\ell$ background, including both its normalization and shape. For the normalization, the PDF used to calculate $R_{W/Z}$ was updated from {\sc MSTW2008 NNLO} to {\sc MSHT20 AN3LO}. For the shape, the default template generated at NLL+NLO accuracy with {\sc CTEQ6.1} was replaced by one generated at NNLL+NNLO accuracy using the {\sc NNPDF3.1 nnlo} PDF set. The combined impact of these updates on the extracted $W$ boson mass is summarized in Table \ref{tab:overall}.

	\begin{table}
	\centering
	\begin{tabular}{p{3.6cm}|p{3.6cm}|l|rrr|rrr}
		\hline
		PDF for $R_{W/Z}$ & PDF & Model & $m_{T}^{e}$ & $p_{T}^{e}$ & $p_{T}^{\nu}$ & $m_{T}^{\mu}$ & $p_{T}^{\mu}$ & $p_{T}^{\nu}$ \\
		\hline
		MSTW2008 nnlo & CTEQ6.1 & NLL+NLO & 0 & 0 & 0 & 0 & 0 & 0 \\
		\hline
		MSHT20 an3lo & NNPDF3.1 nnlo & NNLL+NNLO & -0.3 & -0.3 & -0.3 & -5.5 & -5.8 & -8.6 \\
		\hline
	\end{tabular}
	\caption{The differences between the measured $m_W$ values and the input $m_W$ value, $\Delta m_W = m_W^{(\text{obs})}- m_W^{(\text{in})}$, with the different shapes of the $Z\rightarrow \ell \ell$ background predicted from different PDFs and different calculations. Additional $p_T^Z$ reweighting is applied} 
	\label{tab:overall}
\end{table}

From Table \ref{tab:overall}, the overall impact is estimated to be a shift of -5.5~MeV on the muon channel \mW{} measured value from the \mT{} distribution,  -5.8~MeV from the \pTl{} distribution and -8.6~MeV from the \pTnu{} distribution, which is significantly larger than the uncertainty quoted in Ref.~\cite{CDF2TeV}.

In addition to the impact arising from the modeling of the $Z\rightarrow \ell\ell$ background, there are also technical aspects that could contribute to inaccuracies in the background estimation. However, these details are not clearly addressed in the original paper. One such aspect concerns the cross-section ratio, $R_{W/Z}$, which depends on the mass windows used in the calculation of the total cross sections. For the signal process $W\rightarrow \ell\nu$, the mass window is relatively wide; in our calculation, it is taken as $20 < m_{\ell\nu} < 1000$\,GeV. For the background $Z\rightarrow \ell\ell$ process, the mass window is much narrower. According to Ref.\cite{MSTW2008}, from which the CDF collaboration adopted the value $R_{W/Z} = 10.96$, the range is $66 < m_{\ell\ell} < 116$~GeV.

This discrepancy in mass windows introduces a subtle but important issue when calculating the acceptance $A_Z$ using Eq.~\ref{eq:acc}. 
\begin{equation}
	A_{Z} = \frac{N_{det}}{N_{par}} = \frac{N_{det}^{20-66} + N_{det}^{66-116} + N_{det}^{116-1000}}{ N_{par}^{66-116}} = \frac{N_{det}^{20-1000}}{N_{det}^{66-116}} \frac{N_{det}^{66-116}}{N_{par}^{66-116}}
	\label{eq:acc}
\end{equation}
Here, $N_{par}$ is the number of events at the particle level and $N_{det}$ is the number of events at the detector level. To ensure correct normalization, $N_{par}$ should be counted within the same mass window used to compute $\sigma_Z$, which is $N_{par}^{66-116}$, the number of events at the particle level within the mass window from 66~GeV to 116~GeV. However, $N_{det}$ should ideally be integrated over a broader mass range, since events with $m_{\ell\ell} < 66$~GeV or $m_{\ell\ell} > 116$~GeV can still contribute to the background after detector smearing. $N_{det}^{<66}$ represents the number of events at the detector level with $m_{\ell\ell} < 66$~GeV, $N_{det}^{66-116}$ represents the number of events at the detector level within the mass window from 66~GeV to 116~GeV and $N_{det}^{>116}$ represents the number of events at the detector level with $m_{\ell\ell} >116$~GeV. If $N_{det}^{<66}$ and $N_{det}^{>116}$ are neglected, the $Z\rightarrow \ell\ell$ background will be underestimated, leading to a bias toward a higher extracted $m_W$ value.

The associated correction factor can be expressed as $N_{det}^{20-1000}/N_{det}^{66-116}$, which is approximately 1.01(2) in the muon channel. This corresponds to an additional shift in the extracted \mW{} of -2.6~MeV for the \mT{} distrinution, -2.1~MeV for the \pTl{} distribution and -3.3~MeV for the \pTnu{} distribution.

%%%%%%%%%%%%%%%%%%%%%%%%%%%%%%%%%%%%%%%%%%%%%%%%%%%%%%%%%%%%%%%%%%%%

\section{Conclusion \label{sec:conclusion}}

The $Z$ boson background plays a critical role in high-precision measurements of the $W$ boson mass, particularly in the muon decay channel. In this study, we investigated the potential impact of mismodeling the $Z$ boson background on the CDF measurement of the $W$ boson mass. Our analysis indicates that such mismodeling could introduce a bias of up to 8 MeV in the muon channel, shifting the measured $W$ boson mass closer to the Standard Model (SM) prediction. While this potential effect is significantly larger than the uncertainty on the $Z$ boson background modeling reported by CDF, it is not sufficient to fully explain the observed discrepancy of over 4$\sigma$ between the CDF measurement and the SM prediction, or the tension with all other $m_W$ measurements.

To improve the robustness and reproducibility of the CDF result, it would be useful to study directly all simulated samples used in the analysis once they become publicly available. Such transparency would enable independent cross-checks and allow the broader community to better assess the modeling choices and their impact on the final result. Ultimately, a reanalysis of the CDF data using modern, consistent, and fully simulated templates for both $W$ and $Z$ boson production might provide a more robust evaluation of the potential systematic effects associated with the $Z$ boson background.

%%%%%%%%%%%%%%%%%%%%%%%%%%%%%%%%%%%%%%%%%%%%%%%%%%%%%%%%%%%%%%%%%%%%

\section*{Acknowledgement}

Chen Wang gratefully acknowledges the support of the Alexander von Humboldt Foundation through a postdoctoral scholarship, under which parts of this work were carried out.

\bibliographystyle{unsrt}
\bibliography{./Bibliography}

\begin{thebibliography}{10}

\bibitem{gfitter}
J.~Haller, A.~Hoecker, R.~Kogler, K.~M{\"o}nig, T.~Peiffer, J.~Stelzer, and
  The~Gfitter Group.
\newblock Update of the global electroweak fit and constraints on
  two-higgs-doublet models.
\newblock {\em The European Physical Journal C}, 78(8):675, 2018.

\bibitem{Isaacson:2022rts}
Joshua Isaacson, Yao Fu, and C.~P. Yuan.
\newblock {resbos2 and the CDF W mass measurement}.
\newblock {\em Phys. Rev. D}, 110(9):094023, 2024.

\bibitem{LHC-TeVMWWorkingGroup:2023zkn}
Simone Amoroso et~al.
\newblock {Compatibility and combination of world W-boson mass measurements}.
\newblock {\em Eur. Phys. J. C}, 84(5):451, 2024.

\bibitem{Zhang:2024hzr}
Rui Zhang and Zhen Zhang.
\newblock {Double Parton Scattering Effect on the Measurement of $W$-Boson
  Mass}.
\newblock 11 2024.

\bibitem{CDF:2012gpf}
T.~Aaltonen et~al.
\newblock {Precise measurement of the $W$-boson mass with the CDF II detector}.
\newblock {\em Phys. Rev. Lett.}, 108:151803, 2012.

\bibitem{D0:2012kms}
Victor~Mukhamedovich Abazov et~al.
\newblock {Measurement of the W Boson Mass with the D0 Detector}.
\newblock {\em Phys. Rev. Lett.}, 108:151804, 2012.

\bibitem{CDF2TeV}
T.~Aaltonen et~al.
\newblock {High-precision measurement of the $W$ boson mass with the CDF II
  detector}.
\newblock {\em Science}, 376(6589):170--176, 2022.

\bibitem{CDF:2013dpa}
Timo~Antero Aaltonen et~al.
\newblock {Combination of CDF and D0 $W$-Boson Mass Measurements}.
\newblock {\em Phys. Rev. D}, 88(5):052018, 2013.

\bibitem{ATLAS7TeV2}
Georges Aad et~al.
\newblock {Measurement of the W-boson mass and width with the ATLAS detector
  using proton\textendash{}proton collisions at $\sqrt{s}=7$ TeV}.
\newblock {\em Eur. Phys. J. C}, 84(12):1309, 2024.

\bibitem{CMS13TeV}
CMS Collaboration.
\newblock High-precision measurement of the w boson mass with the cms
  experiment at the lhc, 2024.

\bibitem{LHCb13TeV}
Roel Aaij et~al.
\newblock {Measurement of the W boson mass}.
\newblock {\em JHEP}, 01:036, 2022.

\bibitem{ATLAS7TeV}
Morad Aaboud et~al.
\newblock {Measurement of the $W$-boson mass in pp collisions at $\sqrt{s}=7$
  TeV with the ATLAS detector}.
\newblock {\em Eur. Phys. J. C}, 78(2):110, 2018.
\newblock [Erratum: Eur.Phys.J.C 78, 898 (2018)].

\bibitem{Powheg}
Paolo Nason.
\newblock A new method for combining nlo qcd with shower monte carlo
  algorithms.
\newblock {\em Journal of High Energy Physics}, 2004(11):040, dec 2004.

\bibitem{Powheg2}
Stefano Frixione, Paolo Nason, and Carlo Oleari.
\newblock Matching nlo qcd computations with parton shower simulations: the
  powheg method.
\newblock {\em Journal of High Energy Physics}, 2007(11):070, nov 2007.

\bibitem{Powheg3}
Simone Alioli, Paolo Nason, Carlo Oleari, and Emanuele Re.
\newblock A general framework for implementing nlo calculations in shower monte
  carlo programs: the powheg box.
\newblock {\em Journal of High Energy Physics}, 2010(6):43, 2010.

\bibitem{CT10}
Hung-Liang Lai, Marco Guzzi, Joey Huston, Zhao Li, Pavel~M. Nadolsky, Jon
  Pumplin, and C.-P. Yuan.
\newblock New parton distributions for collider physics.
\newblock {\em Phys. Rev. D}, 82:074024, Oct 2010.

\bibitem{Geant4}
S.~Agostinelli et~al.
\newblock Geant4—a simulation toolkit.
\newblock {\em Nuclear Instruments and Methods in Physics Research Section A:
  Accelerators, Spectrometers, Detectors and Associated Equipment},
  506(3):250--303, 2003.

\bibitem{DYTurbo}
Stefano Camarda, Maarten Boonekamp, Giuseppe Bozzi, Stefano Catani, Leandro
  Cieri, Jakub Cuth, Giancarlo Ferrera, Daniel de~Florian, Alexandre Glazov,
  Massimiliano Grazzini, Manuella~G. Vincter, and Matthias Schott.
\newblock Dyturbo: fast predictions for drell--yan processes.
\newblock {\em The European Physical Journal C}, 80(3):251, 2020.

\bibitem{ATLAS:2023fsi}
{Improved W boson Mass Measurement using 7 TeV Proton-Proton Collisions with
  the ATLAS Detector}.
\newblock 2023.

\bibitem{LHCb:2021bjt}
Roel Aaij et~al.
\newblock {Measurement of the W boson mass}.
\newblock {\em JHEP}, 01:036, 2022.

\bibitem{CDF:2022hxs}
T.~Aaltonen et~al.
\newblock {High-precision measurement of the $W$ boson mass with the CDF II
  detector}.
\newblock {\em Science}, 376(6589):170--176, 2022.

\bibitem{Martin:2009iq}
A.~D. Martin, W.~J. Stirling, R.~S. Thorne, and G.~Watt.
\newblock {Parton distributions for the LHC}.
\newblock {\em Eur. Phys. J. C}, 63:189--285, 2009.

\bibitem{Harland-Lang:2014zoa}
L.~A. Harland-Lang, A.~D. Martin, P.~Motylinski, and R.~S. Thorne.
\newblock {Parton distributions in the LHC era: MMHT 2014 PDFs}.
\newblock {\em Eur. Phys. J. C}, 75(5):204, 2015.

\bibitem{Bailey:2020ooq}
S.~Bailey, T.~Cridge, L.~A. Harland-Lang, A.~D. Martin, and R.~S. Thorne.
\newblock {Parton distributions from LHC, HERA, Tevatron and fixed target data:
  MSHT20 PDFs}.
\newblock {\em Eur. Phys. J. C}, 81(4):341, 2021.

\bibitem{Ball:2012cx}
Richard~D. Ball et~al.
\newblock {Parton distributions with LHC data}.
\newblock {\em Nucl. Phys. B}, 867:244--289, 2013.

\bibitem{NNPDF:2017mvq}
Richard~D. Ball et~al.
\newblock {Parton distributions from high-precision collider data}.
\newblock {\em Eur. Phys. J. C}, 77(10):663, 2017.

\bibitem{NNPDF:2021njg}
Richard~D. Ball et~al.
\newblock {The path to proton structure at 1\% accuracy}.
\newblock {\em Eur. Phys. J. C}, 82(5):428, 2022.

\bibitem{Stump:2003yu}
Daniel Stump, Joey Huston, Jon Pumplin, Wu-Ki Tung, H.~L. Lai, Steve Kuhlmann,
  and J.~F. Owens.
\newblock {Inclusive jet production, parton distributions, and the search for
  new physics}.
\newblock {\em JHEP}, 10:046, 2003.

\bibitem{Lai:2010vv}
Hung-Liang Lai, Marco Guzzi, Joey Huston, Zhao Li, Pavel~M. Nadolsky, Jon
  Pumplin, and C.~P. Yuan.
\newblock {New parton distributions for collider physics}.
\newblock {\em Phys. Rev. D}, 82:074024, 2010.

\bibitem{Dulat:2015mca}
Sayipjamal Dulat, Tie-Jiun Hou, Jun Gao, Marco Guzzi, Joey Huston, Pavel
  Nadolsky, Jon Pumplin, Carl Schmidt, Daniel Stump, and C.~P. Yuan.
\newblock {New parton distribution functions from a global analysis of quantum
  chromodynamics}.
\newblock {\em Phys. Rev. D}, 93(3):033006, 2016.

\bibitem{Hou:2019efy}
Tie-Jiun Hou et~al.
\newblock {New CTEQ global analysis of quantum chromodynamics with
  high-precision data from the LHC}.
\newblock {\em Phys. Rev. D}, 103(1):014013, 2021.

\bibitem{Minnlo}
Pier~Francesco Monni, Paolo Nason, Emanuele Re, Marius Wiesemann, and Giulia
  Zanderighi.
\newblock Minnlops: a new method to match nnlo qcd to parton showers.
\newblock {\em Journal of High Energy Physics}, 2020(5):143, 2020.

\bibitem{Minnlo2}
Pier~Francesco Monni, Emanuele Re, and Marius Wiesemann.
\newblock {MiNNLO$_{\text {PS}}$: optimizing $2\rightarrow 1$ hadronic
  processes}.
\newblock {\em Eur. Phys. J. C}, 80(11):1075, 2020.

\bibitem{Sjostrand:2014zea}
Torbj\"orn Sj\"ostrand, Stefan Ask, Jesper~R. Christiansen, Richard Corke,
  Nishita Desai, Philip Ilten, Stephen Mrenna, Stefan Prestel, Christine~O.
  Rasmussen, and Peter~Z. Skands.
\newblock {An introduction to PYTHIA 8.2}.
\newblock {\em Comput. Phys. Commun.}, 191:159--177, 2015.

\bibitem{Photos}
P.~Golonka and Z.~Was.
\newblock Photos monte carlo: a precision tool for qed correctionsin z and w
  decays.
\newblock {\em The European Physical Journal C - Particles and Fields},
  45(1):97--107, 2006.

\bibitem{Photos2}
N.~Davidson, T.~Przedzinski, and Z.~Was.
\newblock Photos interface in c++: Technical and physics documentation.
\newblock {\em Computer Physics Communications}, 199:86--101, 2016.

\bibitem{CT18Z}
Tie-Jiun Hou, Jun Gao, T.~J. Hobbs, Keping Xie, Sayipjamal Dulat, Marco Guzzi,
  Joey Huston, Pavel Nadolsky, Jon Pumplin, Carl Schmidt, Ibrahim Sitiwaldi,
  Daniel Stump, and C.-P. Yuan.
\newblock New cteq global analysis of quantum chromodynamics with
  high-precision data from the lhc.
\newblock {\em Phys. Rev. D}, 103:014013, Jan 2021.

\bibitem{Resbos}
C.~Bal\'azs and C.-P. Yuan.
\newblock Soft gluon effects on lepton pairs at hadron colliders.
\newblock {\em Phys. Rev. D}, 56:5558--5583, Nov 1997.

\bibitem{CTEQ6M}
Jonathan Pumplin, Daniel~Robert Stump, Joey Huston, Hung-Liang Lai, Pavel
  Nadolsky, and Wu-Ki Tung.
\newblock New generation of parton distributions with uncertainties from global
  qcd analysis.
\newblock {\em Journal of High Energy Physics}, 2002(07):012, aug 2002.

\bibitem{ResBos2}
Joshua~Paul Isaacson.
\newblock {\em {ResBos2 : precision resummation for the LHC ERA}}.
\newblock PhD thesis, Michigan State U., Michigan State U., 2017.

\bibitem{Nadolsky:2008zw}
Pavel~M. Nadolsky, Hung-Liang Lai, Qing-Hong Cao, Joey Huston, Jon Pumplin,
  Daniel Stump, Wu-Ki Tung, and C.~P. Yuan.
\newblock {Implications of CTEQ global analysis for collider observables}.
\newblock {\em Phys. Rev. D}, 78:013004, 2008.

\bibitem{RunITune}
F.~Landry, R.~Brock, P.~M. Nadolsky, and C.-P. Yuan.
\newblock Fermilab tevatron run-1 z boson data and the collins-soper-sterman
  resummation formalism.
\newblock {\em Phys. Rev. D}, 67:073016, Apr 2003.

\bibitem{Melnikov:2006kv}
Kirill Melnikov and Frank Petriello.
\newblock {Electroweak gauge boson production at hadron colliders through
  $O(\alpha_s^2)$}.
\newblock {\em Phys. Rev. D}, 74:114017, 2006.

\bibitem{Gavin:2010az}
Ryan Gavin, Ye~Li, Frank Petriello, and Seth Quackenbush.
\newblock {FEWZ 2.0: A code for hadronic Z production at
  next-to-next-to-leading order}.
\newblock {\em Comput. Phys. Commun.}, 182:2388--2403, 2011.

\bibitem{MSTW2008}
A.~D. Martin, W.~J. Stirling, R.~S. Thorne, and G.~Watt.
\newblock Parton distributions for the lhc.
\newblock {\em The European Physical Journal C}, 63(2):189--285, 2009.

\end{thebibliography}
	
\end{document}